\documentclass[12pt]{iopart}

\usepackage{iopams}
\usepackage{amssymb}
\usepackage{graphicx}

{\catcode`\|=\active
  \gdef\Braket#1{\left<\mathcode`\|"8000\let|\BraVert {#1}\right>}}
\def\BraVert{\egroup\,\vrule\,\bgroup}
{\catcode`\|=\active
  \gdef\set#1{\mathinner{\lbrace\,{\mathcode`\|"8000\let|\midvert #1}\,\rbrace}}
  \gdef\Set#1{\left\{\:{\mathcode`\|"8000\let|\SetVert #1}\:\right\}}}
\def\midvert{\egroup\mid\bgroup}
\def\SetVert{\egroup\;\vrule\;\bgroup}

\begin{document}

\title[Steady-state properties of coupled Ising chains]{Steady-state properties of coupled hot and cold Ising chains}

\author{Maxim O Lavrentovich}

\address{Physics Department, Harvard University, 17 Oxford Street, Cambridge, MA 02138, U.S.A.
\ead{mlavrent@physics.harvard.edu}}
\begin{abstract}
Recently, the author and Zia (2010) reported on exact results for  a far-from-equilibrium system in which two coupled semi-infinite Ising chains at temperatures $T_h$ and $T_c$, with $T_h>T_c$, establish a flux of energy across their junction.  This paper provides a complete derivation of those results, more explicit expressions for the energy flux, and a more detailed characterization of the system at arbitrary $T_c$ and $T_h$.   We consider the two-point correlation functions and the energy flux $F(x)$ between each spin, located at integer position $x$, and its associated heat bath.   In the $T_h \rightarrow \infty$ limit, the flux $F(x)$ decays exponentially into the cold bath 
(spins with $x=1,2,\ldots$) for all $T_c>0$ and transitions into a power law decay as $T_c \rightarrow 0$.  We find an asymptotic expansion for large $x$ in terms of modified Bessel functions that captures both of these behaviors.  We perform Monte Carlo simulations  that give excellent agreement with both the exact and asymptotic results for $F(x)$.  The simulations are also used to study the system at arbitrary $T_h$ and $T_c$.
\end{abstract}

\maketitle

\section{Introduction}

           The study of non-equilibrium systems has countless applications in many areas, including physics, biology, chemistry, and economics.  As discussed in a recent review of non-equilibrium statistical mechanics \cite{ZiaR}, the formulation of a universal description of non-equilibrium systems, analogous to the Gibbs framework for equilibium systems, has been widely acknowledged as an important goal.  Simple models, like the kinetic Ising model, provide us with important tools to build  intuition and to provide test cases for more general treatments.   Studies of such simple systems, reviewed in \cite{ZiaR, ZiaBook, privman}, have greatly contributed to our understanding of non-equilibrium phenomena.     One-dimensional models are particularly interesting as they are often amenable to analytic methods.  Moreover, low-dimensional models can  be easily simulated on a computer, providing us with another tool for exploring non-equilibrium systems.   Finally, kinetic Ising models are particularly useful as they can be mapped to models of other dynamics (e.g., of particles and surfaces) and can therefore characterize a broad class of non-equilibrium phenomena with possible experimental realizations \cite{privman}. 

We will be interested in Ising models  driven out of equilibrium via couplings to  heat baths with different temperatures.  These heat baths set up temperature gradients which induce macroscopic energy fluxes in the system.     
  Given the simplicity of the Ising model, it is sometimes possible to choose couplings such that analytic results for the energy fluxes  \cite{RaczZia94}, steady-state correlation functions, corrections to the Boltzmann distribution \cite{SS, SS2}, and even the full time-dependent behavior \cite{MZS} of the system are available.  In most previous studies of this kind, the systems have a translational symmetry (see \cite{ZiaBook} for a review).  For example, a well-studied Ising chain model has alternating spins  coupled to two heat baths with different temperatures.  Infinite range models with multiple temperatures have also been considered where all of the spins interact \cite{InfR}.  Finally, there have been studies (see \cite{Quantum1, Quantum2, Quantum3}) of quantum spin chains where an energy flux is induced directly by an applied field.

An arguably more realistic way to drive an Ising spin chain out of equilibrium is by linking together the ends of two sub-chains, each held at a different temperature.  This breaks the translational invariance of the system and leads to nontrivial spatial profiles for quantities such as the energy flux.  Such a localized jump in temperature is found in many systems, such as  at the interface between the air and a space heater.   In the context of kinetic Ising models, a possible experimental realization was suggested by Schm\"{u}ser  and Schmittmann \cite{SS2}: Nuclei in a crystalline solid can be prepared at a particular spin temperature.  Two adjacent domains of nuclei at different spin temperatures might be a way to realize the kind of         system considered here.  Also, some exact analytic results are already available for such  systems in one dimension (see \cite{farago1,farago2} and discussion in \cite{MLRKPZ2010}) where one spin  injects energy into an Ising chain via random flipping.  In two dimensions, models using Kawasaki dynamics (where neighboring spin states are exchanged)  reveal  more interesting features of such driven systems, such as convection cells \cite{convection}.  In this paper, we find exact results for a one-dimensional model where a localized temperature gradient establishes energy fluxes through the system with interesting spatial properties.    

 In a recent paper  \cite{MLRKPZ2010}, the author and Zia presented exact results for an infinite kinetic Ising chain with its left half coupled to a hot heat bath and the its right half coupled to a colder bath. A sharp temperature gradient is established at the junction between the two halves, driving the system far from equilibrium.
  In this paper, we will give a complete account of the brief analysis presented in \cite{MLRKPZ2010}.  We calculate the exact expressions for the steady-state two-spin correlation functions and use them to compute the flux of energy $F(x)$ between a spin at integer location $x$ and its associated heat bath.  Intuitively, $F(x)$   describes how energy flows from the hot to cold bath due to the temperature gradient. We discuss both the case where the two baths are at arbitrary temperatures and the limit in which the hot bath approaches an infinite temperature.   In the latter limit, we find excellent approximations to the asymptotic behavior of $F(x)$ for spins in the cold bath that are far from the junction.   

In section~\ref{sec:setup} we establish our notation and the microscopic details of the model. In section~\ref{sec:arbtemp} we express the energy fluxes $F(x)$ in terms of the two-point correlation functions, which we calculate for arbitrary temperatures for the hot and cold baths. As these expressions are rather unwieldy, in section~\ref{sec:inftemp} we calculate $F(x)$ exactly for the case where the hot bath temperatures goes to infinity.  For large $x$, we find that $F(x)$ decays exponentially with $x$ for cold bath temperatures $T_c > 0$.  As $T_c \rightarrow 0$, $F(x)$  decays as a power law, with $F(x) \sim x^{-3}$.  We find excellent approximations for $F(x)$ capturing both of these large $x$ behaviors in section~\ref{sec:asymp}.  All of the exact and approximate results  for the infinitely hot heat bath limit are affirmed by Monte Carlo simulations in section~\ref{sec:simul}. We also use the simulations to explore the behavior of $F(x)$ for arbitrary hot and cold bath temperatures.   We present possible outlooks for future studies and make concluding remarks in section~\ref{sec:final}.

\section{The Model \label{sec:setup}}

We consider a kinetic Ising chain in one dimension with $2N$ spins.  For consistency, our notation will be very similar to the notation used  in the previous paper \cite{MLRKPZ2010}.  The variables $\sigma_x = \pm 1$ will denote the two possible values of the spin at site $x = 0, \pm 1, \pm 2,\ldots,\pm N$.  Although there are many ways to interpret the two values, here we will use the language of magnetic spins so that the values $\pm 1$ will denote a spin pointing up or down, respectively.  We begin our analysis by defining a Hamiltonian $\mathcal{H} ( \{ \sigma_x \})$ for each configuration of Ising spins.  For the case of a chain with nearest-neighbor interactions with a ferromagnetic constant coupling constant $J > 0$, the Hamiltonian is
\begin{equation}
\mathcal{H}(\{ \sigma_x \} ) =- \sum_{\langle x,y \rangle} J \, \sigma_x \, \sigma_{y} =- \sum_{x=-N}^{N-1} J \sigma_x \sigma_{x+1} , \label{eq:Hamiltonian}
\end{equation}
where we sum over all nearest neighbor pairs $\langle x, y \rangle$ in the chain.  For a chain at equilibrium with a single heat bath at temperature $T$, we can compute a canonical distribution function over spin configurations: $P_{\mathrm{eq}}( \{ \sigma_x \}) = Z^{-1} e^{- \beta \mathcal{H} (\{\sigma_x \})}$, where $\beta \equiv (k_B T)^{-1}$,  and $Z$ is the normalization factor or partition function. 

 We will be interested in driving our system out of equilibrium by coupling the spins to heat baths of different temperatures.  Unlike the equilibrium case, we have to specify the particular coupling between the heat baths and the spins.  The standard way to do this is to have the spins transition from one configuration $\{\sigma_x\}$ to another $\{\sigma_x'\}$ with probability rates $W( \{ \sigma_x \} \rightarrow \{ \sigma_x' \})$.   In the equilibrium case, where all the spins are coupled to one temperature, these rates are chosen to satisfy the detailed balance condition $W(\{\sigma_x\} \rightarrow \{\sigma_x'\})P_{\mathrm{eq}}(\{\sigma_x\},t)=W(\{\sigma_x'\} \rightarrow \{\sigma_x\})P_{\mathrm{eq}}(\{\sigma_x'\},t)$.  In the spirit of the corresponding equilibrium studies, we choose the simple Glauber spin-flip dynamics, originally formulated for the equilibrium case \cite{glauber}.  In these dynamics, the two configurations $\{\sigma_x\}$ and $\{\sigma_x '\}$ differ by a single spin flip.  The rates $W$ for flipping the spins will depend on the temperature of the bath to which the chain is coupled. We will generalize these Glauber rates to include a location-dependent heat bath  temperature $T(x)$ (with its inverse $\beta(x)$).   Then, we consider the probability rates $w_x(\sigma_x) \equiv w_x (\sigma_x \rightarrow -\sigma_x)$ of flipping a spin at location $x$ (with $-N < x < N$) given by
  \begin{equation}
 w_x(\sigma_x) =\frac{1}{2\Delta t}  \left[ 1 - \frac{\gamma(x)}{2} \, \sigma_x ( \sigma_{x-1} + \sigma_{x+1})\right]
, \label{eq:MErates}
\end{equation}
where $\gamma(x) \equiv \tanh [ 2  \beta(x) J]$ and $\beta(x)=[k_B T(x)]^{-1}$, where $T(x)$ is the temperature of the heat bath coupled to spin $x$. The time step $\Delta t$ sets the time scale at which the spin flipping occurs. For simplicity, we will choose this scale so that $\Delta t = 1$. Since $J > 0$ (so $\gamma > 0$), if the spin at $x$ is anti-aligned  with its neighbors, then the flipping rate is proportional to $1+\gamma(x)>1$ and to $1-\gamma(x)<1$  if it is aligned.  Thus, as expected for a ferromagnet, the spins tend to align as  $\gamma$ increases (i.e., temperature decreases).  

We now have to specify what to do at the chain boundaries at $x = \pm N$.   One choice is to set $\sigma_N = \sigma_{-N}$ and employ periodic boundary conditions with the same flipping rate as in equation (\ref{eq:MErates}).  Another choice is to allow for open boundary conditions in which the boundary spin flipping rates  are chosen to satisfy the detailed balance condition in the equilibrium case, yielding $w_{\pm N}(\sigma_{\pm N}) =[1- \omega(\pm N) \sigma_{\pm N} \sigma _{\pm N \mp 1}]/(2 \Delta t )$, where $\omega(x) \equiv \tanh [   \beta(x) J] $ \cite{OpenIsing}.   In the following, we will mostly ignore these boundary conditions as we are interested in the behavior of the system as $N \rightarrow \infty$  where we assume the effects of the boundaries will be negligible.   In a completely rigourous treatment, this assumption needs to be checked.  In this paper we will check this  by comparing our analytic results with simulations, which will use the open boundary conditions described above.     Finally, we will set $J=1$ so that all of our energies will be given in units of $J$.

 When $\beta(x)$ is not a constant, the spins can no longer achieve thermal equilibrium as the heat baths at each spin $x$ will compete with each other to create a temperature gradient, establishing thermal energy fluxes in the system which persist in the steady-state.  In this case, the detailed balance condition is broken and our steady state distribution  (which we assume the system goes into as $t \rightarrow \infty)$ is governed by the time-independent probability distribution
\begin{equation}
P_*(\{\sigma_x\}) = P(\{\sigma_x\},t \rightarrow \infty),
\end{equation}
 which is different from the equilibrium distribution $P_{\mathrm{eq}}(\{\sigma_x\})$. The standard technique to deal with this kind of system is to analyze the master equation for the probability distribution $P(\{\sigma_x \},t)$ of observing a configuration $\{ \sigma_x \}$  of spins at time $t$.  For the non-equilibrium Glauber dynamics we consider, it is given by
\begin{equation}
\partial_t P(\{\sigma_x\},t)  =  \frac{1}{N_{\mathrm{tot}}} \sum_{q=-N}^N \left[w_q(-\sigma_q )P(\{\sigma_x  \}_q,t) -w_q(\sigma_q ) P(\{\sigma_x\},t)\right],  \label{eq:MEquation}
\end{equation}
where the total number of spins is $N_{\mathrm{tot}} = 2N+1$ and $\{ \sigma_x \}_q$ is identical to $\{\sigma_x\}$, with the exception of a single flip of the $q$-th spin.   The two terms in the summation in equation (\ref{eq:MEquation}) represent the change in $P(\{\sigma_x\},t)$ due to transitions into and out of the configuration $\{\sigma_x\}$ via spin flipping at spin $q$ at rate $w_q(\sigma_q)$.   In this paper, we will be interested in the steady-state solution $ P_*(\{\sigma_x\}) \equiv P(\{\sigma_x\},t \rightarrow \infty)$ of equation (\ref{eq:MEquation}), for which the left-hand side is equal to zero. 

\begin{figure}[!ht]
\centering
\includegraphics[width=5in]{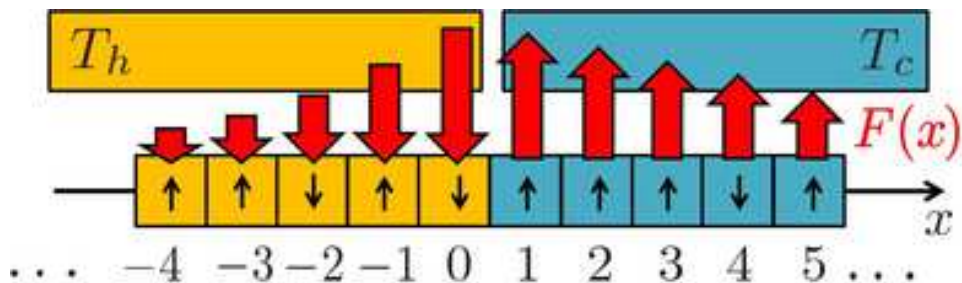}
\caption{\label{fig:setup}  A schematic  of two adjacent kinetic Ising spin chains coupled to heat baths at different temperatures $T_h$  and $T_c$, with $T_h > T_c$.  The red arrows show the flow of heat $F(x)$ between the heat baths and the chains.   Heat flows across the junction from the hot bath and into the cold one. As  $|x|$ increases, the heat flow will decrease as the spins approach equilibrium with their respective heat baths away from the junction. This is illustrated by the decreasing size of the red arrows.  We expect that the heat flux decays faster in the hot bath, as discussed in the main text (see section \ref{sec:simul}).}
\end{figure}

To fully specify our model, we now   consider a localized ``junction'' at spin $x = 0$ between two Ising chains at different temperatures, as illustrated in figure \ref{fig:setup}.   The left chain's heat bath will be set to a temperature $T(x)=T_h$ for $x = -N,-N+1,\ldots,0$, while the right bath will be set to $T(x)=T_c$ for $x=1,2,\ldots,N$, such that $T_h > T_c$.  We define corresponding $\gamma(x)$  parameters which will satisfy
\begin{equation}
\gamma(x) = \cases
{
\gamma_h & for $-N \leq x \leq 0$ \\
\gamma_c & for $0 < x \leq N$
} \label{eq:gammadef}
\end{equation}
and $\gamma_c > \gamma_h$.
Intuitively, we expect the temperature gradient to induce a flow of heat from the hot bath into the cold bath.  The most dramatic non-equilibrium behavior will occur at the junction between the chains at location $x = 0$.  Far away from this junction, i.e. for $x \rightarrow \pm \infty$ in the thermodynamic limit ($N \rightarrow \infty$),  we expect that the chains are at equilibrium with their respective baths and  the heat flow decays to zero.  An illustration of the model and the expected qualitative behavior are given in figure~\ref{fig:setup}.   Our objective now is to compute how this energy flux $F(x)$ decays away from the junction as a function of the spin location $x$ in the stationary state characterized by $P_*(\{\sigma_x\})$.    
\section{Two Heat Baths at Arbitrary Temperatures \label{sec:arbtemp}}

In order to find the net energy flux $F(x)$ at each spin, we recognize that all contributions to $F(x)$ will come from spin flips at spin $x$.  Then, given our Hamiltonian in (\ref{eq:Hamiltonian}), we see that  the energy change $\Delta E(x)$ \textit{gained} by the heat bath (or \textit{lost} by the chain) due to a single spin flip at location $x$ is given in units of $J$ as
\begin{equation}
 \Delta E(x) = \mathcal{H}(\{ \sigma_z \} ) -\mathcal{H}(\{ \sigma_z \}_x )=-2\sigma_x(\sigma_{x-1} + \sigma_{x+1}),
\end{equation} 
where $\sigma_x$ is the value of the spin at $x$ \textit{before} the flip. We also  know the flip rate $\omega_x(\sigma_x)$ at  spin $x$.  Therefore, the net heat flux $F(x)$ must be given by  \cite{RaczZia94}
\begin{eqnarray}
F(x)  & = & \left\langle \omega_x(\sigma_x) \Delta E(x) \right\rangle \nonumber \\[5pt]
& =&  \gamma(x)(1+\left\langle \sigma_{x-1} \sigma_{x+1} \right\rangle) - \left\langle \sigma_x \sigma_{x+1} \right\rangle - \left\langle \sigma_x  \sigma_{x-1} \right\rangle. \label{eq:generaleflux}
\end{eqnarray}
    As we are interested in the average solution in the stationary state, the bracket averages $\langle \ldots \rangle$ are with respect to the stationary distribution $P_*(\{\sigma_x\})$.  We can think about $F(x)$  as a flow of heat from the spin chain \textit{into} the heat bath.  We have $F(x)<0$ for heat flowing into the chain and $F(x)>0$ for flow out of the chain. 

We see that to compute $F(x)$, we have to consider the two point correlation functions 
\begin{equation}
 \left\langle \sigma_x \sigma_y \right\rangle = \sum_{\{\sigma_z\}} \sigma_x \sigma_y \, P_*(\{\sigma_z\}), \label{eq:2pntdef}
\end{equation}
for which we define a convenient notation $\langle x, y \rangle \equiv \langle \sigma_x \sigma_y \rangle$.  Using a calculation completely analogous to the one done by Glauber for the equilibrium case \cite{glauber}, it is possible to derive a difference equation for $\langle x , y \rangle$ from our master equation in (\ref{eq:MEquation}).  First, we set the left-hand side of equation (\ref{eq:MEquation}) to zero to get an equation for $P_*(\{\sigma_z\})$.    Then we multiply this equation by $\sigma_x \sigma_y$ and sum over all configurations $\{\sigma_z\}$.  This gives us an equation for the two-point correlation functions $\langle x  , y \rangle$, which, after some  manipulations, we can write as
\begin{eqnarray}
\fl -2\sum_{\{\sigma_z \}}\sigma_x \sigma_y \left[w_x(\sigma_x ) +w_y(\sigma_y )\right] P_*(\{\sigma_z\}) = 0 \nonumber   \\[5pt]
\fl  \gamma(x) \left( \left\langle x+1,y \right\rangle+ \left\langle x-1,y \right\rangle \right) + \gamma(y)\left(\left\langle x,y+1 \right\rangle+ \left\langle x, y-1 \right\rangle  \right) - 4 \left\langle x,y \right\rangle = 0 \nonumber \\[5pt]
\fl \left[\gamma(x)  \delta_x^2 + \gamma(y) \delta_y^2+ 2\left(\gamma(x)+\gamma(y)-2 \right) \right] \left\langle x,y \right\rangle =0, \label{eq:Helmholtz}
\end{eqnarray}
where $\delta_x^2$ is a second order difference operator which acts on any function $f(x)$ as $\delta_x^2 f(x) = f(x+1)+f(x-1)-2f(x)$.  Notice that equation (\ref{eq:Helmholtz}) has the form of an anisotropic discrete Helmholtz equation, which can be solved using Green's function techniques.  To do this, we require appropriate boundary conditions.  Given the definition of the two-point correlation function in (\ref{eq:2pntdef}), we must have the two boundary conditions (BCs)
\begin{equation}
\cases{  \left\langle x,x \right\rangle = 1 \\
\left\langle x,y \right\rangle = \left\langle y,x \right\rangle }. \label{eq:BCs1}
\end{equation}
 Another boundary condition comes from the behavior of the system for large separations $|x-y|$.  Namely, we expect that as $|x-y| \rightarrow \infty$, the spins become uncorrelated so that $\left\langle x,y \right\rangle \rightarrow \left\langle \sigma_x \right\rangle \left\langle \sigma_y \right\rangle=0$,  since there is no spontaneous magnetization in one dimension and  $\left\langle \sigma_x \right\rangle = 0$ for any choice of $0<\gamma(x)\leq 1$  and $x$, as discussed in \cite{MLRKPZ2010}.   Since we are dealing with a finite system for now, we will have to implement this boundary condition by enforcing it exactly at the boundary spins at $x = \pm N$.    So, we will also have the BCs
\begin{equation}
\left\langle x, \pm N \right\rangle = 0 \label{eq:BCs2}
\end{equation}
for all $x \neq \pm N$.

\begin{figure}[!ht]
\centering
\includegraphics[width=4in]{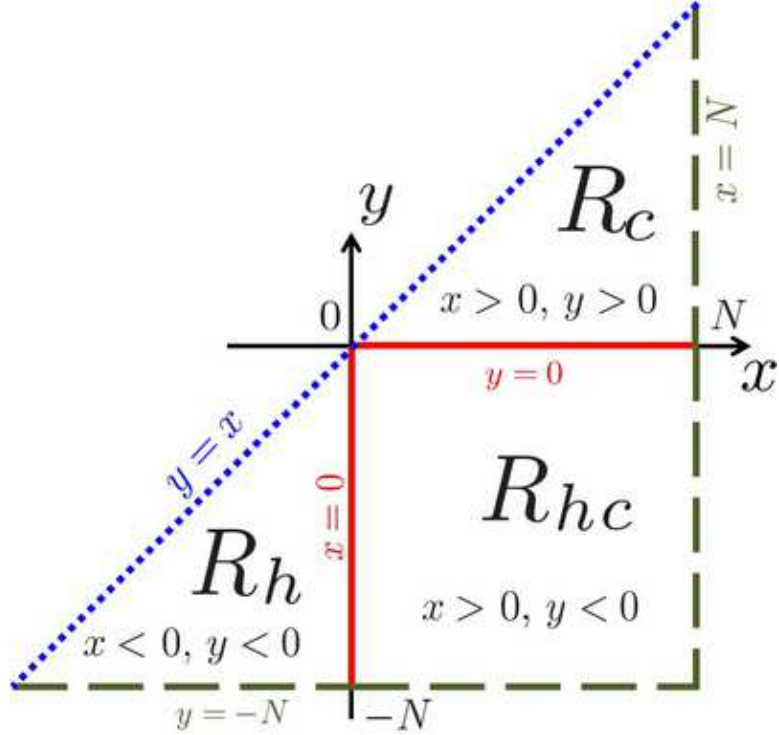}
\caption{\label{fig:BCs}  The regions  $R_h$, $R_{hc}$, and $R_c$ represent correlations between spins coupled to the hot, hot and cold, and cold bath, respectively.  We want to solve the equations in (\ref{eq:regioneqs}) for these correlations.  The dashed green lines ($y=-N$ and $x=N$) correspond to the BCs given in (\ref{eq:BCs2}).  We must enforce the first condition in (\ref{eq:BCs1}) on the dotted blue line $(y=x)$.  Finally, we have to ensure our solutions match along the boundaries between the regions indicated by the solid  red lines ($x=0$ and $y=0$). }
\end{figure}

The second condition in (\ref{eq:BCs1}) implies that we can now just consider $x \geq y$  for the purposes of calculating the correlation functions.   Then, given our definition of $\gamma(x)$ in (\ref{eq:gammadef}), we can identify three regions in the $(x,y)$ plane on which we must solve equation (\ref{eq:Helmholtz}) with the appropriate BCs. These regions are illustrated in figure \ref{fig:BCs}.  Substituting the appropriate values for $\gamma(x)$ from (\ref{eq:gammadef}) into equation (\ref{eq:Helmholtz}), we find the equations
\begin{equation}
\fl \cases{\left[  \gamma_h\delta_x^2 +\gamma_h  \delta_y^2+ 4\left(\gamma_h-1 \right) \right] \left\langle x,y\right\rangle =0 & in  $R_h$: $y <  0$, $x <0$ \\
\left[\gamma_c  \delta_x^2 + \gamma_h \delta_y^2+ 2\left(\gamma_c+\gamma_h-2 \right) \right] \left\langle x,y \right\rangle =0& in  $R_{hc}$: $y < 0$,  $x > 0$ \\
\left[ \gamma_c\delta_x^2 +  \gamma_c\delta_y^2+ 4\left(\gamma_c-1  \right) \right] \left\langle x,y \right\rangle =0& in $R_{c}$: $y > 0$, $x > 0$}, \label{eq:regioneqs}
\end{equation}
the solutions to which must be matched along the red lines $x = 0$ and $y = 0$ shown in figure \ref{fig:BCs}.

To find the solutions to equation (\ref{eq:Helmholtz}) in the various regions, we  look at the Green's function $G(x,y \, ; \, \xi, \eta)$ in each region $R_{h,hc,c}$  that satisfies 
\begin{eqnarray}
\fl \mathcal{D}_{x,y} G(x,y \, ; \, \xi,\eta) & \equiv & \left[\gamma(x)  \delta_x^2 + \gamma(y) \delta_y^2+ 2\left(\gamma(x)+\gamma(y)-2 \right) \right] G(x,y \, ; \, \xi, \eta) \nonumber \\
& = &\delta_{x,\xi} \, \delta_{y,\eta} \label{eq:Greenseq}
\end{eqnarray}
where $\gamma(x)$ is given in (\ref{eq:gammadef}),  $\delta_{x,y}$ are Kronecker delta functions, and $\mathcal{D}_{x,y}$ is a convenient notation for the discrete operator in the square brackets. Our Green's function is chosen to satisfy $G(x,y \, ; \, \xi, \eta) = 0$ along the  red, green, and blue lines in figure \ref{eq:Greenseq}.   Performing a Fourier transform, we find a solution for $G(x,y \, ; \, \xi, \eta)$ for values of   $(x,y)$ and $(\xi , \eta)$ in the three regions of interest:
\begin{equation}
\fl G(x,y  ; \xi, \eta) = \cases{  \sum_{m=-N}^{0} \sum_{n=-N}^{m}  \frac{U_{m,n}(x,y) U_{m,n}(\xi,\eta)}{4-2\gamma_h[\cos(\pi m/N)+\cos (\pi n/N) ]}  & in $R_h$ \\
 \sum_{m=0}^{N} \sum_{n=-N}^{0} \frac{V_{m,n}(x,y)V_{m,n}(\xi,\eta)}{4-2\gamma_c\cos(\pi m/N)-2 \gamma_h\cos (\pi n/N) }& in  $R_{hc}$ \\
 \sum_{m=1}^{N} \sum_{n=1}^{m} \frac{U_{m,n}(x,y) U_{m,n}(\xi,\eta)}{4-2\gamma_c[\cos(\pi m/N)+\cos (\pi n/N) ]} & in $R_c$ }, \label{eq:greensexp}
\end{equation}
where we take the  linear combinations of  Fourier eigenfunctions  which vanish on the region boundaries and are normalized over the  regions:
\begin{equation}
\fl\cases{
U_{m,n}(x,y) =\frac{2}{N}  \left[ \sin \left( \frac{\pi m x}{N} \right) \sin \left( \frac{\pi n y}{N} \right)-\sin \left( \frac{\pi my}{N} \right) \sin \left( \frac{\pi n x}{N} \right) \right] \\
V_{m,n}(x,y)=\frac{2}{N} \sin \left( \frac{\pi m x}{N} \right) \sin \left( \frac{\pi n y}{N} \right)
}. \label{eq:eigenfuncs}
\end{equation}
Using this Green's function, we are now able to solve for the two-point correlation functions.

 We find the correlation function $\left\langle x,y \right\rangle$ by exploiting a discrete Green's theorem.  Namely, if we multiply equation (\ref{eq:Greenseq}) by $\left\langle \xi,\eta \right\rangle$ and sum over all values of $\xi$ and $\eta$ in the region we are interested in, we find that 
\begin{equation}
\left\langle x,y \right\rangle = \sum_{(\xi,\eta) \in R_{h,hc,c}} \left[ \left\langle \xi, \eta  \right\rangle \mathcal{D}_{\xi,\eta}G(x,y \, ;  \xi, \eta)-G(x, y \, ;  \xi , \eta) \,\mathcal{D}_{\xi,\eta} \left\langle \xi, \eta  \right\rangle  \right], \label{eq:Greens2nd}
\end{equation}
where we  recognize that $\mathcal{D}_{\xi,\eta} \left\langle \xi, \eta \right\rangle = 0$.   One can show that the summations over all $(\xi,\eta) \in R_{h,hc,c}$ in (\ref{eq:Greens2nd}) reduce to summations just over the boundaries $\partial R_{h,hc,c}$ of each region.  In particular, for the three regions we are interested in, we have the solutions (using the more compact notation $G_{x,y}(\xi,\eta) \equiv G(x,y \, ;  \xi ,\eta)$)
\begin{equation}
\fl \left\langle x,y \right\rangle = \cases{  \gamma_h\sum_{\eta =-N}^{0} \left[G_{x,y}(-1 ,\eta )\left\langle 0,\eta
\right\rangle + G_{x,y}(\eta+1 ,\eta) +G_{x,y}(\eta,\eta-1 )\right] & in $R_h$ \\
\gamma_{c}\sum_{\eta =-N}^{0}G_{x,y}(1,\eta)\left\langle0,\eta\right\rangle +\gamma_{h}\sum_{\xi
=1}^{N}G_{x,y}(\xi , -1)\left\langle \xi ,0 \right\rangle& in $R_{hc}$ \\
\gamma_{c}\sum_{\xi =1}^{N} \left[G_{x,y}(\xi ,1)\left\langle \xi ,0\right\rangle + G_{x,y}(\xi ,\xi -1) +G_{x,y}(\xi +1,\xi) \right]& in $R_c$ }, \label{eq:corrsolns}
\end{equation}
where  we have simplified some boundary terms by applying the boundary conditions (Eqs.~\ref{eq:BCs1}, \ref{eq:BCs2}).    Finally, we must find the functions $\left\langle \xi, 0 \right\rangle$ and $\left\langle 0, \eta \right\rangle$ that match the solutions in the three different regions and are consistent with Eq~\ref{eq:Helmholtz}.  
 We now have a complete description of the general solution, but finding an analytical form for $\langle x,y \rangle$ is cumbersome.  To make progress, we will now look at the energy flux through the chain in the case that the hot bath is infinitely hot.  This limit will allow us to get a much more detailed understanding of the behavior of this model.

\section{ Hot Heat Bath at Infinite Temperature \label{sec:inftemp}}

Let us now set $\gamma  \equiv \gamma_c$, $T \equiv T_c$, etc.  We then  set  $\gamma_h = 0$ to keep the hot bath at infinite temperature.  We immediately see from (\ref{eq:regioneqs}) that all of the correlations in region $R_h$ vanish. In region $R_{hc}$ all the correlations vanish, except for the ones at the boundary: $\langle x, 0 \rangle$ for $x > 0$.  This is intuitive because, for $T_h \rightarrow \infty$, all of the spins in the hot bath flip randomly.  Thus, any spin in that bath is equally likely to point up or down and cannot be correlated with a non-neighboring spin in the cold bath.  Explicitly,  we see from (\ref{eq:regioneqs}) that for $\gamma_h = 0$, the correlation function $\langle x, y \rangle$ satisfies
\begin{equation}
\left[\gamma  \delta_x^2 + 2\left(\gamma-2 \right) \right] \left\langle x,y \right\rangle = \gamma \left(\langle x-1 ,y \rangle+\langle x+1,y \rangle \right)  -4 \langle x,y \rangle  
=  0 \label{eq:singletempeq}
\end{equation}
for all $y < 0$.  This is a homogeneous, linear second order difference equation which we can easily solve using standard techniques. Namely, we try to find the solution of the form $\langle x, y \rangle = A  r^x$.  Substituting this ansatz into equation (\ref{eq:singletempeq}) and solving for $r$, we find that there are two possible values of $r$ so that the general solution for $\langle x, y \rangle$ is given by
\begin{equation}
\langle x, y \rangle = A_1 \tilde{\omega}^x + A_2 \tilde{\omega}^{-x}, \label{eq:diffeqansatz}
\end{equation}
where $A_{1,2}$ are constants set by the BCs and 
\begin{equation}
 \tilde{\omega} \equiv \frac{2}{\gamma} - \sqrt{ \frac{4}{\gamma^2} -1}.
\end{equation}
Then, since all of the correlations in $R_h$ vanish,  we know  that $\langle 0, y \rangle = 0$ for all $y = -N+1,\ldots,-1$.  This means that we must have $A_1 = - A_2 \equiv B/2$ so that our general result in $R_{hc}$ becomes $\langle x , y \rangle = B \sinh (x \ln \tilde{\omega} )$.   We now let $N \rightarrow \infty$ and require that that $\langle x , y \rangle \rightarrow 0$ as $x \rightarrow \infty$ for any $y$.  The only way this can happen is if $B = 0$.  This means that the $\langle x, y \rangle$ vanishes in region $R_{hc}$, as well.
The energy fluxes for $x < 0$  must also vanish since we will have, via (\ref{eq:generaleflux}),
\begin{equation}
F(x) =-   \left[ \left\langle x,x+1 \right\rangle + \left\langle x,x-1\right\rangle\right] = 0 \qquad \mbox{ for all } x < 0.
\end{equation}

We now compute the correlations in region $R_c$ and the boundary function  $\left\langle x ,0 \right\rangle$ for $x \geq 1$.  The latter function satisfies equation (\ref{eq:singletempeq}) with $y  = 0$.  Therefore, we have the same general solution given in (\ref{eq:diffeqansatz}).  This time, our boundary conditions are  $\langle 0,0 \rangle = 1$ and  $\langle x, 0 \rangle \rightarrow 0$ as $x \rightarrow \infty$.  Since $0<\tilde{\omega}<1$,  these boundary conditions set $A_2 = 0$ and $A_1 = 1$ in (\ref{eq:diffeqansatz}), yielding the solution 
  
\begin{equation}
\langle x, 0 \rangle = \tilde{\omega}^{x} \qquad \mbox{ for } x \geq 0.  \label{eq:boundsoln}
\end{equation}
Consequently, at the $x = 0$ spin, we have an energy flux equal to
\begin{equation}
F(x=0) =- \tilde{\omega}.  \label{eq:fluxat0}
\end{equation}

We have now shown that \textit{all} of the flux from the hot bath into the spin chain is located at the boundary spin at $x = 0$.   The negative sign in the expression for  $F(x=0)$ means that, as expected, the heat flows into the chain from the infinitely hot bath.     With equation (\ref{eq:corrsolns}) in  region $R_c$, we now have a complete solution to our correlation functions $\langle x, y \rangle$ and, consequently, $F(x)$ for all $x > 0$.  However, the resulting expression for $F(x)$ is cumbersome.  To find a simpler solution, we now recognize that the Helmholtz equation in $R_c$ (see (\ref{eq:regioneqs}))  has the same form as for an equilibrium Ising chain (a chain of spins all coupled to a single heat bath with a parameter $\gamma$).  The only difference between the equilibrium solution  $\langle x, y \rangle_{\mathrm{eq}}$ and the non-equilibrium one $\langle x, y \rangle$ must be in the BCs.  The $\langle x, x\rangle = 1$ BC is the same in both cases, so the difference must come from the $\langle x , 0\rangle$ correlation we computed in (\ref{eq:boundsoln}). In the equilibrium case, we can find this function by solving equation (\ref{eq:Helmholtz}), setting $\gamma(x) = \gamma(y) = \gamma$.  We find the standard exponentially decaying result \cite{Pathria}
\begin{equation}
\langle x, y \rangle_{\mathrm{eq}} =  \omega^{|y-x|} = \left[\frac{1}{\gamma} - \sqrt{ \frac{1}{\gamma^2} -1} \right]^{|y-x|} = e^{- |y-x|/ \xi_{\mathrm{eq}
}}, \label{eq:equilibriumcorr}
\end{equation}
where we have the equilibrium correlation length $\xi_{\mathrm{eq}} \equiv -(\ln \omega)^{-1}$ and we have used the relation $\omega = \gamma^{-1} - \sqrt{\gamma^{-2}-1}$ for $\omega = \tanh \beta  $ and $\gamma = \tanh ( 2  \beta )$.  Notice that the equilibrium and non-equilibrium solutions for $\langle  \xi, 0 \rangle$ are related by a replacement of $\gamma$ with $\gamma/2$, the average of the parameters of the hot and cold chain:  $0$ and $\gamma$.   

Given the close relationship between the non-equilibrium and equilibrium cases,  it is convenient to study the deviation $\Delta(x,y) \equiv \langle x,y \rangle - \langle x ,y \rangle_{\mathrm{eq}}$.
Both $\langle x,y \rangle $ and $\langle x, y \rangle_{\mathrm{eq}}$ satisfy  (\ref{eq:regioneqs}) in $R_c$ with $\gamma \equiv \gamma_c$ and  the linearity of the Helmholtz equation allows us to use (\ref{eq:corrsolns}) to solve for $\Delta(x,y)$.  The only difference is in the boundary condition along the $y = 0$ line. There we must have
\begin{equation}
\Delta(x,0) = \tilde{\omega}^x - \omega^x .\label{eq:DeltaBC}
\end{equation}
Thus, the solution to $\Delta (x,y)$ in $R_c$ is
\begin{equation}
\Delta(x,y) =\gamma\sum_{\xi =0}^{N} G(x,y \,;\xi ,1)\left[ \tilde{\omega}^{\xi} - \omega^{\xi}   \right]  \qquad \mbox{ for } 0 < y < x. \label{eq:diffsolns}
\end{equation}
Substituting the expression for $G(x,y\,; \xi,1)$ in equation (\ref{eq:greensexp}) into (\ref{eq:diffsolns}), we find  that 
 \begin{eqnarray}
\fl \Delta(x,y)  &  =  \frac{\gamma}{N}   \sum_{m=1}^{N} \sum_{n=0}^{m-1}   \frac{ U_{m,n}(x,y)}{2-\gamma[\cos(\pi m/N)+\cos (\pi n/N) ]}   \nonumber \\[5pt]
 \fl&   \quad \times  \left[ \sum_{\xi =0}^{N} \left( \tilde{\omega}^{\xi} - \omega^{\xi}   \right)  \left[\sin \left( \frac{\pi m \xi }{N} \right) \sin \left( \frac{\pi n }{N} \right)-\sin \left( \frac{\pi m}{N} \right) \sin \left( \frac{\pi n \xi}{N} \right) \right]  \right]    \nonumber \\[5pt]
\fl & = \frac{ \gamma}{N}  \sum_{m=1}^{N} \sum_{n=0}^{m-1}  \frac{U_{m,n}(x,y) \sin(\pi m/N)\sin (\pi n/N)}{2-\gamma[\cos(\pi m/N)+\cos (\pi n/N) ]} \left[ \frac{\tilde{\omega}}{1 - 2\tilde{%
\omega}\cos(\frac{\pi m }{N}) + \tilde{\omega}^2}   \right.  \nonumber \\[5pt]
 \fl &   \left. - \frac{\tilde{\omega}}{1 - 2\tilde{\omega}%
\cos \left(\frac{\pi n }{N} \right) + \tilde{\omega}^2}+ \frac{\omega }{1 - 2 \omega %
\cos  \left(\frac{\pi n }{N} \right) + \omega^2} -  \frac{\omega}{1 - 2
\omega\cos \left (\frac{\pi m }{N} \right) + \omega^2} \right],  \label{eq:diffexact}
\end{eqnarray}
where the sum over $\xi$ was performed by expanding the $\sin( \pi m \xi/N)$, $\sin (\pi n \xi/N)$ terms in exponentials and summing the resultant geometric series.  We have also made the approximation that $\tilde{\omega}^N \approx \omega^N \approx 0$ for large $N$.  This approximation will be exact in the $N \rightarrow \infty$ limit for any $0 < \gamma < 1$.  We now move to the calculation of the flux $F(x)$ for $x > 0$.

Since  $F(x) = 0$ at equilibrium, we can replace $\langle \ldots \rangle$ with $\langle \ldots \rangle_{\mathrm{eq}}$ in (\ref{eq:generaleflux}) to get a combination of equilibrium correlation functions that must equal zero.  Then, we  subtract this combination from the right-hand side of (\ref{eq:generaleflux}) to find  
\begin{equation}
F(x)  = \gamma \Delta(x+1,x-1) -\Delta(x+1,x) - \Delta(x,x-1) . \label{eq:fluxwithdiff}
\end{equation}
Since (\ref{eq:fluxwithdiff}) includes the  functions $\Delta(x,x- 1)$ and $\Delta(x+1,x)$ and we impose the boundary condition in  (\ref{eq:DeltaBC}), we will have to treat the $x = 1$ case separately. For now, suppose $x > 1$.  Then, substituting (\ref{eq:diffexact}) and (\ref{eq:eigenfuncs}) into equation (\ref{eq:fluxwithdiff}), and using the identities $\tilde{\omega} = 2 \gamma^{-1} - \sqrt{4\gamma^{-2}-1}$ and $\omega = \gamma^{-1} - \sqrt{\gamma^{-2}-1}$ , we get
\begin{eqnarray}
&\fl  F(x) = \frac{2 \gamma^2 }{N^2}
\sum_{m=1}^{N} \sum_{n=1}^{m}\left[ \frac{\sin\left(\frac{\pi nx}{N}\right) \cos\left(\frac{\pi mx}{N}\right) \sin\left(\frac{\pi m}{N}\right) \left( 1
- \gamma  \cos\left(\frac{\pi n}{N}\right) \right) }{2 - \gamma \left[\cos\left(\frac{\pi m}{N}\right) +
\cos\left(\frac{\pi n}{N}\right) \right]}- n \leftrightarrow m\right]
  \nonumber  \\[5pt]
&\fl \qquad \qquad  \times \left[ \frac{3-\gamma \cos\left(\frac{\pi n}{N}\right)-\gamma \cos\left(\frac{\pi m}{N}\right)}{ \left[2-\gamma \cos\left(\frac{\pi n}{N}\right) \right] \left[1-\gamma \cos\left(\frac{\pi m}{N}\right) \right]} -  n \leftrightarrow m  \right]\sin\left(\frac{\pi m}{N}\right) \sin\left(\frac{\pi n}{N}\right), \label{eq:fluxdiscreteexact}
\end{eqnarray}
where the $n \leftrightarrow m$ term is just the first term in each bracket with $n$ and $m$ exchanged. Moving to the $N \rightarrow \infty$ limit, we define the continuous variables
\begin{equation}
k = \frac{\pi m}{N}\qquad \mbox{and} \qquad p = \frac{\pi n}{N}.
\end{equation}
The summations in (\ref{eq:fluxdiscreteexact}) become integrals as we can identify them as regular Riemann sums with step size $\pi /N$. Also, notice that the summand in (\ref{eq:fluxdiscreteexact}) is symmetric under the exchange of $m$ and $n$, vanishes for $m = n$, and is even in both $m$ and $n$.  So, moving to the $N \rightarrow \infty$ limit, making use of the symmetry arguments, and simplifying some of the terms gives us
\begin{eqnarray}
\fl F(x) = \mbox{Re} \left\{ \frac{ \gamma^2}{4i\pi^2}
\int_{-\pi}^{\pi} \int_{-\pi}^{\pi} \mathrm{d}k \, \mathrm{d}p \,e^{i(p+k)x} \left[ \frac{ \sin{p } \left( 1 - \gamma \, \cos{k} \right)-\sin{k } \left( 1 - \gamma \, \cos{p} \right)}{2 -
\gamma (\cos{k} + \cos{p})}   \right] \right.\nonumber \\[5pt]
\fl \qquad \qquad   \left. \times\left[ \frac{1}{2-\gamma \cos{k}} -
\frac{1}{2-\gamma \cos{p}} + \frac{1}{1 - \gamma \cos{p}} -
\frac{1}{1 - \gamma \cos{k}}\right] \sin{k} \sin{p}\right\}, \label{eq:pkintegral}
\end{eqnarray}
where we have written our expression as the real part of a complex function that will aid us in what follows.

\begin{figure}[!ht]
\centering
\includegraphics[width=5in]{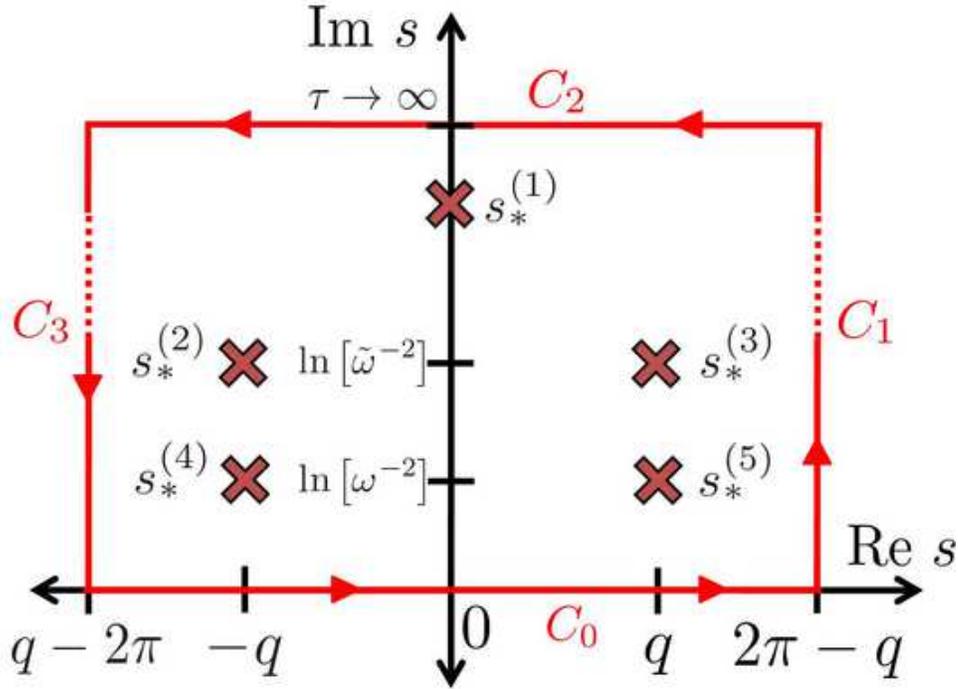}
\caption{\label{fig:contour}  Here we denote the contour $\mathcal{C} = C_0+C_1+C_2+C_3$ in the complex plane that we require to perform the integration over $s$ in (\ref{eq:sqintegral2}).  We will let $\tau \rightarrow \infty$ so that the contour becomes a semi-infinite strip.  The red crosses denote the locations of the poles that are picked up by the contour. The locations of the  poles $s_*^{(i)}$ ($i=1,\ldots,5$) picked up by $\mathcal{C}$. }
\end{figure}

To make progress calculating the integral in (\ref{eq:pkintegral}), it is convenient to change variables to $s = k+p$ and $q=k-p$. Since our region of integration is $k \in (-\pi,\pi)$ and $p \in (-\pi , \pi)$, we can let $q$ range from $(-2 \pi, 2 \pi)$ and have $s \in (|q|-2 \pi, 2\pi - |q|)$. Changing to these variables and applying some trigonometric identities yields
\begin{equation}
 F(x) = \mbox{Re} \left\{ \int_0^{2\pi} \int_{q-2 \pi }^{2 \pi -q} H_x(s,q) \, \mathrm{d}s \, \mathrm{d} q \right\},   \label{eq:sqintegral2}
\end{equation}
where  
\begin{eqnarray}
\fl H_x(s,q)  =   \frac{ i \gamma^3e^{isx}}{\pi^2} \,  \left[ \frac{\left[ \cos ^2\left( \frac q2\right) -\cos
^2\left( \frac s2\right) \right]  \, \left[  \gamma \cos\left( \frac{q}{2} \right) - \cos\left(\frac{s}{2}\right)\right]}{1 - \gamma \cos \left( \frac s2\right) \cos \left( \frac q2\right) }  \right]
  \nonumber    \\[5pt]
 \fl  \qquad \qquad \qquad \qquad \qquad \qquad  \times   \sum_{a=1}^2  \frac{(-1)^a \sin \left( \frac s2\right) \sin^2 \left( \frac q2\right) }{2 a^2-4 \gamma a \cos \left( \frac s2\right) \cos \left( \frac q2\right)  + \gamma^2(\cos s+\cos q)}  .\label{eq:gdef} 
\end{eqnarray}
We have also recognized that the real part of the integrand in (\ref{eq:sqintegral2}) is even in  $q$ so that we can change our domain of integration to $q \in (0,2 \pi)$ and $s \in (2\pi -q,q-2 \pi)$.  We first compute the integral over $s$ in (\ref{eq:sqintegral2}) via a rectangular contour integration in the complex $s$ plane.  The contour  $\mathcal{C}$ is shown in figure \ref{fig:contour}.  Cauchy's residue theorem tells us that our integral of interest (the one over $C_0$ in figure \ref{fig:contour}) is given by
\begin{eqnarray}
 \int_{q-2 \pi }^{2 \pi - q}\mathrm{d} s\, H_x(s,q)  = 2\pi i \sum_{j=1}^5  \,\mbox{Res}\left[s^{(j)}_* \right]- \int_{C_1} H_x(s,q) \, \mathrm{d} s \nonumber \\
 \qquad \qquad \qquad \qquad  \qquad \qquad - \int_{C_2} H_x(s,q) \, \mathrm{d} s - \int_{C_3} H_x(s,q) \, \mathrm{d} s, \label{eq:gsintegral}
\end{eqnarray}
where the first term represents the contribution from the poles inside $\mathcal{C}$ located at $s_*^{(j)}$, with residues  $\mbox{Res}[s_*^{(j)}]$ .  We shall see in the following that there will be five poles.   Before we move on to these poles, consider the integrals over the various parts of the contour $\mathcal{C}$.  First,  the contribution from $C_2$ vanishes
since the factor $e^{i s x} = e^{itx-\tau x}$ will go exponentially to zero as $\tau \rightarrow \infty$.  The integral over $C_1$ is a bit trickier to analyze and is given by
\begin{equation}
\int_{C_1} \mathrm{d} s \, H_x(s,q) = i \int_0^{\infty} \mathrm{d} t \, H_x(2\pi-q+it,q) \equiv \int_0^{\infty} \mathrm{d} t \, h_{C_1}(t,q).
\end{equation}
Using (\ref{eq:gdef}), we can show that the real part of $h_{C_1}(q,t)$ satisfies, for integers $x > 0$,
\begin{equation}
\mbox{Re}\left[h_{C_1}(t,\pi+ q') \right] = -\mbox{Re}\left[h_{C_1}(t,\pi- q') \right],
\end{equation}
for all $ q' \in (-\pi, \pi)$ and $t > 0$.    Thus, the contribution from $C_1$ to $F(x)$ must vanish since we will have
\begin{equation}
\mbox{Re} \left\{ \int_0^{2\pi} \int_{C_1} H(s,q) \, \mathrm{d}s \, \mathrm{d} q \right\} =\int_{0}^{\infty}  \int_{-\pi}^{\pi}  \mbox{Re}\left[h_{C_1}(t,\pi+ q') \right]\, \mathrm{d} q'  \, \mathrm{d}t  = 0. 
\end{equation}
The contribution from $C_3$ vanishes by an analogous argument.  All that remains  is to evaluate the contributions from the poles.

Notice that the functions in the numerators and denominators in the expression for  $H(s,q)$ in (\ref{eq:gdef}) are all analytic over the entire complex $s$ plane.  Thus, the poles will come from the zeros of the denominators, i.e., the solutions to the equations 
 \begin{equation}
\cases{1 - \gamma \cos \left( \frac{s}{2} \right) \cos \left( \frac{q}{2} \right) = 0  \\
 2 a^2-4 \gamma a \cos \left( \frac{s}{2} \right) \cos \left( \frac{q}{2} \right)  + \gamma^2(\cos s+\cos q)= 0 \qquad a= 1,2}. \label{eq:zeros}
\end{equation}
   There are many possible such solutions, but only five will be picked up by the contour $\mathcal{C}$, as shown in figure \ref{fig:contour}.  These five   poles are  located at $s_*^{(i)}$ ($i = 1,\ldots,5$), given by 
\begin{equation}
 \cases{s_*^{(1)} = i \,2 \ln \left[ \gamma^{-1}  \sec \left(\frac{q}{2} \right) + \sqrt{ \gamma^{-2} \, \sec^2 \left( \frac{q}{2} \right) - 1} \, \right] 
\\
s_*^{(2,3)} = \pm q-i\, 2  \ln \omega  \\
s_*^{(4,5)} = \pm q - i\,2  \ln \tilde{\omega}  },
\end{equation}
These poles are only picked up by the contour  $\mathcal{C}$  when $0 < q < \pi$:   the real parts of $s_*^{(i)}$ for $i=2,3,4,5$ will not be between $q-2\pi$ and $2 \pi - q$ when $q \geq \pi$, and $s_*^{(1)}$ is pushed off to $+ i \infty$ as $q \rightarrow \pi$.  So, we now can change our limits of integration for the integral over $q$:
\begin{eqnarray}
 F(x) & =-   2 \pi \sum_{
j=1}^5 \int_0^{\pi} \mbox{Im} \left[ \mbox{Res} \left( s_*^{(j)} \right)  \right]     \mathrm{d} q . \label{eq:fluxfromresidue}
\end{eqnarray}
We now compute the residues at each pole.  We deal with the $i=2,3$ poles first.  The residues are, after much simplification,
\begin{equation}
 \mbox{Res} \left[ s_*^{(2,3)} \right] = \frac{i e^{\pm i q (x+1)} \omega ^{2 x} \left(\omega ^2-1\right) \left(\omega ^2-e^{ \mp 2 i q}\right)}{4 \pi ^2 \left(\omega +\omega ^3\right)} ,
\end{equation}
with the upper (lower) sign corresponding to the residue at $s_*^{(2)}$ ($s_*^{(3)}$). Picking up the imaginary part  and performing the integrations yields the same result for both poles (as we expect since $\mbox{Re}[G(s,q)]$ is odd in the $s$ variable):
\begin{eqnarray}
 \int_0^{\pi} \mbox{Im} \left\{ \mbox{Res} \left[ s_*^{(2,3)} \right] \right\} \, \mathrm{d} q & = & \frac{\omega ^{2 x-1} \left(\omega ^2-1\right) \left[1+x+\omega ^2(1-x)\right] \sin[\pi  x]}{4 \pi ^2 \left(x^2-1\right) \left(1 +\omega ^2\right)} \nonumber \\[5pt]
& = &\frac{\omega(1 -\omega ^2)}{4 \pi (1+\omega ^2)}\, \delta_{x1}=\frac{\gamma(1 -\omega ^2)}{8 \pi}\, \delta_{x1}, \label{eq:s23integral}
\end{eqnarray}
where we see that when $x$ is an integer, this contribution vanishes for all integer $x > 1$ and has a finite contribution  at $x = 1$.  Since we are dealing with the $x > 1$ case, we can ignore this contribution.  It will, however, turn out to be useful later.  

 We find a similar situation for the $i=4,5$ poles, with slightly more complicated integrals over $q$.  Here, the residues are
\begin{equation}
\fl \mbox{Res} \left[ s_*^{(4,5)} \right] = \frac{i e^{\pm i q (x+1)}\left(e^{\mp i q}-1\right) \tilde{\omega} ^{2 x} \left(\tilde{\omega} ^2-e^{\mp2iq} \right) \left[e^{ \mp i q}(1-3 \tilde{\omega}^2)+\tilde{\omega} ^2  \left(\tilde{\omega} ^2-3\right)\right]}{4 \pi ^2 \left( e^{\mp 2 i q}+ \tilde{\omega} ^2\right) \tilde{\omega}\left(1 +\tilde{\omega} ^2\right)} \label{eq:res45exact}, 
\end{equation}
where the upper (lower) sign refers to the residue at $s_*^{(4)}$ ($s_*^{(5)}$). To perform the integral of this contribution of $q$, we first expand (\ref{eq:res45exact}) out in  terms proportional to $e^{i q c}$ for some constant $c$.  For the integral of $\mbox{Res} [s_*^{(4)}]$, we find that all of the terms involve    integrals of the form
\begin{equation}
I_{\tilde{\omega}} (\zeta) \equiv \int_0^{\pi} \mathrm{d} q\, \frac{e^{i q \zeta}}{e^{ -2i q}+ \tilde{\omega}^2} =i\tilde{\omega}^{-2}  \int_{S^+} \mathrm{d}z \, \frac{z^{\zeta+1}}{z^2+\tilde{\omega}^{-2}} ,
\end{equation}
where $\zeta \geq - 2$  is a constant and we have made the variable substitution $z \equiv e^{iq}$ so that our new path of integration is $S^+$, the unit semi-circle in the top half of the complex $z$ plane, traversed clockwise.  There are no poles between $S^+$ and the real  axis  because the poles are located at $z_* = \pm i \tilde{\omega}^{-2}$ and we know that $\tilde{\omega}^{-2} > 1$ for all $0 < \gamma \leq 1$.  Thus, deforming the contour over $S^+$ onto the real axis yields 
\begin{eqnarray}
I_{\tilde{\omega}}(\zeta) &  = &i  \int_{-1}^1 \mathrm{d}z \, \frac{z^{\zeta+1}}{1+\tilde{\omega}^{2}z^2}  =  \frac{i}{2}\,(1+(-1)^{\zeta+1} ) \int_0^1 \mathrm{d}u \, \frac{u^{\zeta/2}}{1+ \tilde{\omega}^2 u}\nonumber  \\[5pt]
& = & i[1+(-1)^{\zeta+1} ] \tilde{F}_{\tilde{\omega}}(\zeta), \label{eq:I1integral}
\end{eqnarray}  
where we have made the substitution $u \equiv z^2$, defined
\begin{equation}
\tilde{F}_{\tilde{\omega}}(\zeta) \equiv   \frac{1}{\zeta+2}\, F\left( 1,  \frac{\zeta}{2}+1 ,\frac{\zeta}{2}+2; - \tilde{\omega}^2 \right),
\end{equation}
  and recognized in (\ref{eq:I1integral}) a standard integral representation for the hypergeometric function $F(\alpha,\beta, \gamma ; z)$  (see e.g., equation 3.197 3  in   \cite{Ryzhik}). 

 Using (\ref{eq:I1integral}) to integrate $\mbox{Res} [ s_*^{(4)} ]$, we find that
\begin{eqnarray}
\fl \int_0^{\pi} \mathrm{d} q  \, \mbox{Res} \left[ s_*^{(4)} \right]  =   \frac{\tilde{\omega}^{2x}( e^{i \pi x}+1)}{4 \pi^2 \tilde{\omega}(1+\tilde{\omega}^2)} \left[ \left(  1-3\tilde{\omega}^2\right) \tilde{F}_{\tilde{\omega}}(x-3)+2 \tilde{\omega}^2 \left( \tilde{\omega}^2+ 1 \right) \tilde{F}_{\tilde{\omega}}(x-1)\right. \nonumber  \\[5pt]
\fl \qquad \qquad  \qquad \qquad \qquad \qquad \qquad \qquad \qquad \qquad  \left.  + \tilde{\omega}^4\left(\tilde{\omega}^2  -3  \right) \tilde{F}_{\tilde{\omega}}(x+1)\right] \nonumber  \\[5pt]
\fl \qquad \qquad \qquad \qquad+\frac{\tilde{\omega}^{2x}( e^{i \pi x}-1)\left(  1-\tilde{\omega}^4\right)}{4 \pi^2 \tilde{\omega}(1+\tilde{\omega}^2)} \left[  \tilde{F}_{\tilde{\omega}}(x-2)- \tilde{\omega}^2 \tilde{F}_{\tilde{\omega}}(x)\right] . \label{eq:sstar4integral}
\end{eqnarray}
The  integral we encounter for the $\mbox{Res}[s_*^{(5)}]$ term is very similar.    However, we do not have to calculate anything for the residue at $s_*^{(5)}$ because we can recognize from (\ref{eq:res45exact})  that the imaginary parts of the residues at $s_*^{(4)}$ and $s_*^{(5)}$ are the same.   Thus, we just have to examine (\ref{eq:sstar4integral}) to find the contribution from both residues.  We find  that for integer $x > 1$, these integrals give us purely real contributions.   The only imaginary contribution will come from the limit $x \rightarrow 1$, where the  first term on the right-hand side of equation (\ref{eq:sstar4integral}) picks up a factor of $-i \pi$ since $(e^{i \pi x} + 1)/(x-1) \rightarrow -i \pi$ when $x \rightarrow 1$.   Therefore,
\begin{equation}
 \int_0^{\pi} \mbox{Im} \left\{ \mbox{Res} \left[ s_*^{(4,5)} \right] \right\} \, \mathrm{d} q  = \frac{\tilde{\omega} \left(  3\tilde{\omega}^2-1\right)}{4 \pi (1+\tilde{\omega}^2)}\, \delta_{x1}=\frac{\gamma\left(  3\tilde{\omega}^2-1\right)}{16 \pi}\, \delta_{x1}. \label{eq:s45integral}
\end{equation}

We now conclude that the only contribution to $F(x)$ for $x > 1$ must come from the  $s_*^{(1)}$ residue.  The residue turns out to be pure imaginary and substituting it into equation (\ref{eq:fluxfromresidue}) gives us 
\begin{equation}
F(x)   =  \frac{4}{\pi \gamma} \int_0^1 \frac{ \left[(\gamma \eta)^{-1}-\sqrt{(\gamma \eta)^{-2}-1}\right]^{2 x} \left(1-\gamma ^2 \eta ^4\right)}{   \left[1+\gamma ^2 \eta ^2 \left(\eta ^2-1\right)\right] \sqrt{1-\eta^2}}\, \mathrm{d} \eta  \quad \mbox{ for }x > 1, \label{eq:finalfexact0}
\end{equation}
where we changed variables to $\eta \equiv \cos(q/2)$.  Let us finally deal with the $x = 1$ value of the flux. 
We substitute (\ref{eq:DeltaBC}) into equation (\ref{eq:fluxwithdiff}) to find that
\begin{equation}
F(x=1) = \gamma( \tilde{\omega}^2 - \omega^2)-\tilde{\omega}+\omega - \Delta ( 2,1 ).
\end{equation} 
   Notice that the $x \rightarrow 1$ limit of the expression for $F(x)$ in (\ref{eq:fluxdiscreteexact}) coincides with the equation for $-\Delta(2,1)$ in (\ref{eq:diffexact}). So, we are now able to  use the contributions we found in equations (\ref{eq:s23integral}) and (\ref{eq:s45integral}) to calculate $\Delta (2,1)$.  We find
\begin{eqnarray}
\fl F(x=1) & = \gamma( \tilde{\omega}^2 - \omega^2)-\tilde{\omega}+\omega+\frac{4}{\pi \gamma} \int_0^1 \frac{ \left[(\gamma \eta)^{-1}-\sqrt{(\gamma \eta)^{-2}-1}\right]^{2 } \left(1-\gamma ^2 \eta ^4\right)}{   \left[1+\gamma ^2 \eta ^2 \left(\eta ^2-1\right)\right] \sqrt{1-\eta^2}}\, \mathrm{d} \eta \nonumber \\[5pt]
 \fl & \qquad \qquad  +\frac{\gamma  \left(  1-3\tilde{\omega}^2\right)}{4}-\frac{\gamma(1 -\omega ^2)}{2}  \nonumber \\[5pt]
\fl & =\frac{4}{\pi \gamma} \int_0^1 \frac{ \left[(\gamma \eta)^{-1}-\sqrt{(\gamma \eta)^{-2}-1}\right]^{2 } \left(1-\gamma ^2 \eta ^4\right)}{   \left[1+\gamma ^2 \eta ^2 \left(\eta ^2-1\right)\right] \sqrt{1-\eta^2}}\, \mathrm{d} \eta .
\end{eqnarray}  
We now see that the behavior of $F(x)$ for \textit{all} $x \geq 1$ is governed by the residue at $s_*^{(1)}$.  This was previously argued for just $x > 1$ in \cite{MLRKPZ2010}, but we have now carefully calculated $F(x)$ for all $x \geq 0$.   In summary, we have the exact expression
\begin{equation}
\fl \qquad \qquad F(x) =\frac{4(1-\delta_{x0})}{\pi \gamma} \int_0^1 \frac{ \left[(\gamma \eta)^{-1}-\sqrt{(\gamma \eta)^{-2}-1}\right]^{2x } \left(1-\gamma ^2 \eta ^4\right)}{   \left[1+\gamma ^2 \eta ^2 \left(\eta ^2-1\right)\right] \sqrt{1-\eta^2}}\, \mathrm{d} \eta  -  \tilde{\omega} \delta_{x0} 
\label{eq:finalfexact}
\end{equation}
for all integers $x \geq 0$.

 The flux of energy must be conserved globally so that the flux into the system from the hot bath ($-\tilde{\omega}$) must be compensated by the total flux out of the cold bath $ \sum_{x>0} F(x)$.  Since the expression in the brackets $[ \ldots ]^{2x}$ in (\ref{eq:finalfexact}) is in the interval $(0,1)$ for any $\gamma,\eta \in(0,1)$, we can perform the summation over all $x$ to find the amusing integral identity
\begin{equation}
\fl \frac{1}{\tilde{\omega}}\sum_{x=1}^{\infty} F(x) =\frac{2}{\pi\gamma \tilde{\omega}}\int_0^1\frac{\left(1-\gamma ^2 \eta ^4\right) \left(\sqrt{1-\gamma ^2 \eta ^2}-1\right)^2}{    \sqrt{1-\eta ^2} \left[1+\gamma ^2 \eta ^2 \left(\eta ^2-1\right)\right] \left(\gamma ^2 \eta ^2+\sqrt{1-\gamma ^2 \eta ^2}-1\right)} =1  
\end{equation}
for all $\gamma \in (0,1)$. This identity can be evaluated numerically as a kind of check of our exact solution.

\section{Energy Flux Asymptotics \label{sec:asymp}}

Looking at the large $x$ behavior of $F(x)$ allows us to characterize the most important physical features of the model.  We shall see that the asymptotics tell us about a cross-over in $F(x)$ from an exponential to a power law decay as $T \rightarrow 0$.   To find these large $x$ asymptotics of $F(x)$, we have to examine the integral in (\ref{eq:finalfexact})
\begin{equation}
\fl I_F (x)\equiv  \int_0^{\pi/2} \frac{ \exp \left\{- 2x \ln \left[[\gamma \cos\theta]^{-1}+\sqrt{\left[\gamma \cos\theta\right]^{-2}-1}\right] \right\} \left[1-\gamma ^2 \cos^4 \theta\right]}{  1-(\gamma/2) ^2 \sin^2(2 \theta)}\, \mathrm{d} \theta, \label{eq:Ifintegral}
\end{equation}
where we have restored the original integration over $q$ and changed variables $ \theta = q/2$.  We know  the main contribution to $I_F$ will come from the minimum of the function 
\begin{equation}
f(\theta) \equiv \ln \left[ \frac{1}{\gamma \cos \theta}+\sqrt{ \frac{1}{\gamma^2 \cos^2 \theta }-1}\right].
\end{equation} 
This minimum occurs at $\theta = 0$.  In order to be able to apply Watson's lemma \cite{Orszag} to our integral,  we change variables to  
\begin{equation}
s \equiv 2f(\theta)+2 \ln \omega, \label{eq:svariable}
\end{equation}
which will satisfy $s \in(0,\infty)$ whenever $\theta \in (0,\pi/2)$.  Inverting equation (\ref{eq:svariable}) and performing this change of variable, equation (\ref{eq:Ifintegral}) becomes
\begin{equation}
 I_F(x)  = e^{2 x\ln \omega} \int_0^{\infty} \mathrm{d} s\, W(s) e^{-s x} \label{eq:Laplaceform}
\end{equation}
where
\begin{equation}
\fl W(s) \equiv \frac{\eta_s^{2}(  1-\gamma ^2 \eta_s^4)}{   2\left[1+\gamma ^2 \eta^2_s  \left( \eta_s^2-1\right)\right] \sqrt{1-\eta_s^2}} \, \left[\sqrt{1-\gamma^2}  \, \cosh \left( \frac{s}{2} \right) + \sinh \left( \frac{s}{2} \right) \right]  \label{eq:Wdef}
\end{equation}
and
\begin{equation}
\eta_s \equiv\left[ \cosh\left( \frac{s}{2} \right) + \sqrt{1-\gamma^2} \sinh \left( \frac{s}{2} \right) \right]^{-1}.
\end{equation}
The expression for $W(s)$ is extremely unwieldy, but we now see that the integral in  (\ref{eq:Laplaceform}) is in an appropriate form for the application of Watson's lemma.  Thus, we only need to worry about the expansion of $W(s)$ around $s = 0$. At $s = 0$, we see from (\ref{eq:Wdef}) that $W(s)$ develops a square root singularity  $s^{-1/2}$ due to the $\sqrt{1-\eta_s^2}$ term, which vanishes like $\sqrt{1-\eta_s^2} = (1-\gamma^2)^{1/4} \, \sqrt{s}\,(1 + \mathcal{O}(s))$ as $s \rightarrow 0$.  If we extract this singularity, the rest of $W(s)$ is analytic in $s$, which we can expand as a Taylor series around $s = 0$ with coefficients $a_n$.  This yields, 
\begin{equation}
\fl  W(s)    = \frac{1}{\sqrt{s}}\sum_{n=0}^{\infty}  a_n s^n  =  \frac{\left(1-\gamma^2\right)^{5/4}}{2 \sqrt{s}} \left[1 -\frac{\left(8 \gamma^4-29\gamma^2+2\right) s}{8\sqrt{1-\gamma ^2}}+\mathcal{O}(s^2)\right]  .\label{eq:Wsexpansion}
\end{equation}
This expansion and subsequent ones were performed with some assistance from the computer algebra system Mathematica 8.0. The expansion in (\ref{eq:Wsexpansion}) becomes invalid in the $\gamma \rightarrow 1$ limit, as the higher order terms $a_n$ in (\ref{eq:Wsexpansion})  become larger than the lower order ones, becoming divergent for $n \geq 3$.  We will deal with this problem later.  For now, we may conclude by Watson's lemma \cite{Orszag}  that  for $1-\gamma$ sufficiently small,  $I_F(x)$ admits an asymptotic expansion given by
\begin{equation}
I_F (x) \sim e^{2x \ln \omega}  \sum_{n=0}^{\infty} \frac{a_n \Gamma( n +1/2)}{x^{ n+1/2}}. \label{eq:Laplaceexpansion} 
\end{equation}
So,  to second  order, the flux is
\begin{equation}
 \fl F(x )  \sim\frac{  \,2\left(1-\gamma^2\right)^{5/4} \omega^{2x}}{ \gamma  \sqrt{ \pi x}}    \left[ 1+\frac{29\gamma^2-8 \gamma^4-2 }{16x\sqrt{1-\gamma ^2}}+ \mathcal{O} \left[ \left(x \sqrt{1-\gamma^2}\right)^{-2}\right]\right]. \label{eq:FLaplaceexpand}
\end{equation}
The $\omega^{2x} = e^{2x \ln \omega}$ term in (\ref{eq:FLaplaceexpand}) tells us that the energy flux $F(x)$  decays exponentially as $e^{-x/\xi}$,  with a decay length equal to
\begin{equation}
\xi = - \frac{1}{2 \ln \omega} = - \frac{1}{2 \ln \left(\gamma^{-1}-\sqrt{\gamma^{-2}-1} \right)}=  \frac{\xi_{\mathrm{eq}}}{2}\, , \label{eq:corrlength}
\end{equation} 
which we recognize as half the equilibrium correlation length $\xi_{\mathrm{eq}}$ (see (\ref{eq:equilibriumcorr})).   Notice that the quantity $x \sqrt{1- \gamma^2}$ has to be \textit{large} for our expansion in (\ref{eq:FLaplaceexpand}) to remain valid.  Since this is impossible for $\gamma \rightarrow 1$ ($T_c \rightarrow 0$), this approach is invalid in this limit, and we must use another method.

To see what happens as the cold bath temperature drops to zero, we first take the $\gamma \rightarrow 1$ limit in the expression for $W(s)$ in (\ref{eq:Wdef}).  We then find a different analytic structure in the $s$ variable with no square root singularities:
\begin{eqnarray}
\left. W(s) \right|_{\gamma \rightarrow 1} & =\frac{2 (3+\cosh s) \sinh \left(\frac{s}{2}\right) \tanh\left(\frac{s}{2}\right)}{7+\cosh(2 s)} & = \sum_{n=0}^{\infty} b_n s^{2n} \\[5pt]
& = \frac{s^2}{4}-\frac{s^4}{24}-\frac{179 s^6}{23040} +\mathcal{O}(s^8).
\end{eqnarray}
Watson's lemma then tells us that
\begin{equation}
F(x) \sim  \frac{4}{\pi } \sum_{n=0}^{\infty} \frac{b_n (2n)!}{x^{1+2n}} =  \frac{2}{\pi x^3}-\frac{4}{\pi x^{5} } + \mathcal{O}(x^{-7}). \label{eq:powerlawexp}
\end{equation}
We see that $F(x)$ transitions into a power-law decay in the $\gamma \rightarrow 0$ limit.  This behavior is reminiscent of an equilibrium system at a  critical point where we find long range correlations.  The particular leading order behavior, $F(x) \sim x^{-3}$, in this non-equilibrium system is interesting and does not seem to have a simple explanation.

To capture both the power law decay of $F(x)$ in the $\gamma \rightarrow 1$ limit and the exponential decay for $\gamma < 1$, we can no longer apply the  analysis utilizing Watson's lemma described above.  Instead, we want to set up an expansion that captures both the  $x \sqrt{1-\gamma^2} \gg 1$ regime, where $F(x)$ exhibits an exponential decay, and the $x \sqrt{1-\gamma^2} \ll 1$ regime, where $F(x)$ has a power law decay.  To do this, we go back to equation (\ref{eq:Ifintegral}) and rewrite it in the modified form
\begin{equation}
\fl  I_F  (x) =  \frac{\omega^{2x} e^{2x \sqrt{1-\gamma^{2}} }}{2}\,  \int_0^{\pi}     f\left[ \cos \left( \frac{q}{2} \right) \right] e^{2x g \left[ \cos(q/2) \right]} \,\exp \left[ - x \sqrt{4(1-\gamma^2)+q^2} \right]\, \mathrm{d} q, \label{eq:IFrewrite}
\end{equation}
where
\begin{equation}
f(y) \equiv  \frac{  1-\gamma^2 y^4}{   1+\gamma ^2 y^2\left(y^2-1\right)\ }, \label{eq:finIF}
\end{equation}
and
\begin{equation}
\fl  g(y) \equiv    \sqrt{(1-\gamma^{2})+ [\mbox{acos}(y)]^2}-\ln \left[(\gamma y)^{-1}+\sqrt{(\gamma y)^{-2}-1} \right]-\sqrt{1-\gamma^{2}}- \ln  \omega. \label{eq:ginIF}
\end{equation}
As in the Laplace method, we now argue that the main contribution to the integral in (\ref{eq:IFrewrite}) will come from the region $q \approx 0$.  So, we now perform the expansion by analogy with the Laplace method by expanding the term  $f(y) e^{2 x g(y)}$ for $y = \cos(q/2)$  in a power series in $q$ and extending our range of integration to $\zeta \in (0,\infty)$. From the expressions in (\ref{eq:finIF}) and (\ref{eq:ginIF}), we see that only the even powers of $q$ will contribute to the expansion. To facilitate this expansion, we first define the constant $\zeta^2 \equiv 4(1-\gamma^{2})$ and calculate the integral 
 \begin{eqnarray}
\fl \int_0^{\infty} q^{2n} \, e^{-  x \sqrt{\zeta^2+q^2}} \,\mathrm{d} q & = \zeta^{2n+1}\int_1^{\infty} z \left(z^2-1 \right)^{n-1/2} \, e^{- x\zeta z} \,\mathrm{d} z \nonumber \\[5pt]
\fl  & = - \frac{\zeta^{2n}}{ \sqrt{\pi}}\, \frac{\partial}{\partial x}\left[ 2^n\left( x\zeta \right)^{-n} \Gamma\left( n + \frac{1}{2} \right)K_{n}( x\zeta) \right] \nonumber \\
\fl & = \frac{(2n-1)!!\,\left(2 \sqrt{1-\gamma^2} \right)^{n+1}}{ x^{n}} \,  K_{n+1}\left(   2x \sqrt{1-\gamma^2}\right) , \label{eq:BesselKintegral}
\end{eqnarray}
where $n \geq 0$ is an integer, $n!! = n(n-2)\ldots$ for any positive integer $n$ (with $0!!=(-1)!!\equiv 1$),  and we have recognized an integral representation of the modified Bessel function (see e.g., equation 3.387 6 in \cite{Ryzhik}). Using (\ref{eq:BesselKintegral}), we may now set up our asymptotic expansion of the integral in (\ref{eq:IFrewrite}) of the form described in \cite{MLRKPZ2010}:
\begin{equation}
 I_F(x) \sim \omega^{2x} e^{  \zeta x} \sum_{n=0}^{\infty} B_n  \, (\zeta x)^{-n}\,  K_{n+1} \left( \zeta x  \right), \label{eq:IFBesselexpand}
\end{equation}
where 
\begin{equation}
B_n =\frac{(2n-1)!!\,[4(1-\gamma^{2})]^{n+1/2}}{2(2n)!}  \, \left. \frac{\partial^{2n}}{\partial q^{2n}} \left[ f\left[ \cos \left( \frac{q}{2} \right) \right] e^{2x g \left[ \cos(q/2) \right]} \right] \right|_{q=0}. \label{eq:Bncoeff}
\end{equation}
These coefficients can be computed exactly for any $n$.   We find that if we keep the first two orders in $1/x$, we have 
\begin{eqnarray}
\fl F(x) &= \frac{4\omega^{2x} (1-\gamma^2)e^{  2x  \sqrt{1-\gamma^{2}}}}{\pi \gamma}   \left[  (1-\gamma^2)^{1/2}K_1 \left(2 x \sqrt{1-\gamma^2} \right) \vphantom{\frac{z^2}{z}}\right. \nonumber \\[5pt]
\fl &  \qquad  \left. +\frac{\gamma ^2 (3-\gamma^2)}{ 2x}\, K_2 \left(2 x \sqrt{1-\gamma^2} \right) - \frac{5(1-\gamma^2)}{16x}K_3 \left(2x \sqrt{1-\gamma^2} \right)+\mathcal{O}\left( x^{-2} \right) \right] . \label{eq:BesselKexpand}
\end{eqnarray} 
One can easily check that the expansion in  (\ref{eq:BesselKexpand}) captures both terms  in  equation (\ref{eq:FLaplaceexpand}) for $\zeta x \gg 1$   and reduces, for $\zeta x \ll 1$, to $2 x^{-3}/ \pi$. We conclude that this Bessel function expansion faithfully captures the behavior of $F(x)$ in both of the regimes of interest.
In the next section, we will compare these asymptotic results with both simulations and the exact expression for $F(x)$ in (\ref{eq:finalfexact}).

\section{Simulations \label{sec:simul}}

We will now verify our solutions for $F(x)$ via a Monte Carlo routine. In this routine, we initialize our Ising chain of $2N+1$ spins at locations $-N \leq x \leq N$ in a random configuration and then evolve it according to the master equation rates.   This is done by   choosing any spin $\sigma_q$ with equal probability and flipping that spin with probability $\Delta t \,w_q(\sigma_q)$.  We then advance the time $t$ by  $t \rightarrow t+\Delta t$ and repeat the process.  We choose open boundary conditions at the ends.  In the $T_h \rightarrow \infty$ limit, the spins coupled to the hot bath flip with probability $1/2$, regardless of the states of their neighbors.  Therefore, we can perform our simulation in this limit on just the $N$ spins coupled to the cold bath and the single infinitely hot spin at $x = 0$. 

\begin{figure}[!ht]
\centering
\includegraphics[width=4in]{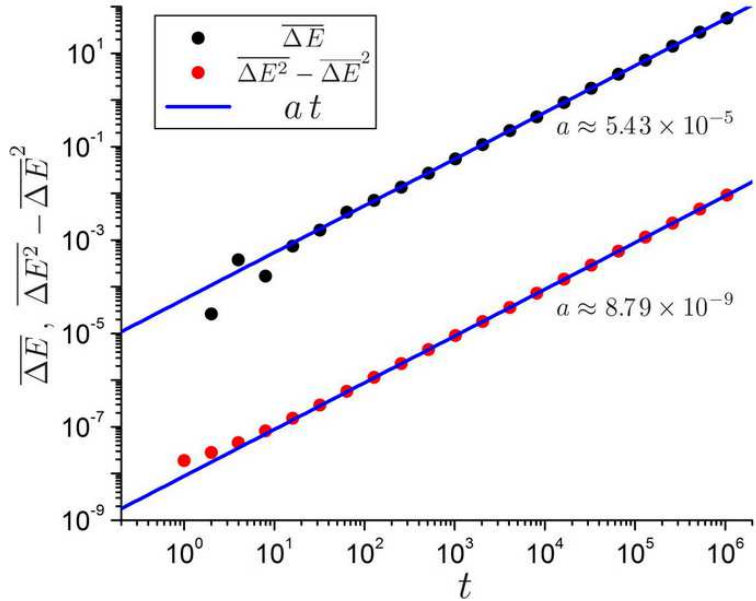}
\caption{\label{fig:Erandomwalk}  A log-log plot of time time dependence of both the mean $\overline{\Delta E(x,t)}$ and the variance $\overline{\Delta E(x,t)^2}-\overline{\Delta E(x,t)}^2$ for $\gamma =0.8$ and $x = 5$. The blue lines show a fit of the data to a linear function and indicate that both the mean and variance grow linearly at large $t$.  The $y$-intercepts of the lines on the log-log plot can be used to estimate the slope $a$ of the fit and, thus,  $\Phi_5$ and $V_5$, as in (\ref{eq:simslopedef}).}
\end{figure}
To record the flux $F(x)$ at each spin, we have to keep track of the spin flips at each spin as each can contribute a change in energy of $\pm4 $, depending on whether the spin is aligned or anti-aligned with its neighbors.  We compute the average change in energy $\overline{\Delta E(x,t)}$ over about  $3 \times 10^{6}$ independent simulation runs after $t = 2^n$ Monte Carlo updates at each spin (for $n=0,1,2,\ldots, 20$).   We tested values of $\gamma$ between $0.5$ and $1$.  For the smaller values of $\gamma$, the sharp exponential decay of $F(x)$ required more simulation runs (up to $1.2 \times 10^7$) to get good statistics at large values of $x$.  We can think about these Monte Carlo updates at each spin as inducing a random walk in the variable $\Delta E(x,t)$, with ``step size'' $4$.   

In an equilibrium case, we expect that the average $\overline{\Delta E(x,t)}$  is zero, as we have no net flux $F(x)$.  However, in our case, the presence of the infinite temperature spin at $x = 0$ induces a \textit{bias} in the $\Delta E(x,t)$ dynamics.  The non-zero value of $F(x)$ induces a uniform drift and, thus, $\overline{\Delta E(x,t)}$ is non-zero and grows linearly in time after the system achieves a steady-state at long times $t$, as shown in the log-log plot in figure \ref{fig:Erandomwalk}.   Notice that the data points  for $\overline{\Delta E(x=5,t)}$ versus  $t$ fall perfectly on a line with unity slope on a log-log plot for large $t$, implying linear growth.  We also see in the figure that the variance, $\overline{\Delta E^2(x,t)}- \overline{\Delta E(x,t)}^2$, also increases linearly in $t$, just as we would expect for a random walk.  In summary, we use these simulation results to estimate the two slopes $V_x$ and $\Phi_x$, where
\begin{equation}
\cases{
\overline{\Delta E(x,t)} \approx \Phi_x t \\
\overline{\Delta E(x,t)^2} - \overline{\Delta E(x,t)}^2 \approx V_x t
}. \label{eq:simslopedef}
\end{equation}
As illustrated by (\ref{eq:generaleflux}), the rate of increase of $\overline{\Delta E(x,t)}$ in the steady-state of the system is equal to the energy flux $F(x)$. Thus, our flux $F(x)$ is estimated by  $F(x) \approx \Phi_x$.  We can also approximate an error in this estimate by utilizing properties of the random walk.    The standard error of the mean $\overline{\Delta E(x,t)}$ is given by
\begin{equation}
\mbox{SE} \left( \overline{\Delta E(x,t)}\right) = \sqrt{ \frac{\overline{\Delta E(x,t)^2} - \overline{\Delta E(x,t)}^2 } { N_{\mathrm{runs}} } }  \label{eq:stderror},
\end{equation}
where $N_{\mathrm{runs}}$ is the number of runs over which we average.  Then, given the definition of $\Phi_x$ in (\ref{eq:simslopedef}), we see from (\ref{eq:stderror}) that an estimate of the error $\pm \Delta F(x)$ in  $F(x) \approx \Phi_x$ is given by
\begin{equation}
\Delta F(x) \approx \sqrt{ \frac{V_x}{N_{\mathrm{runs}}\,t} }\,,
\end{equation}  
where $t$ will be the number of time steps used to determine our slopes $\Phi_x$ and $V_x$.

\begin{figure}[!ht]
\centering
\includegraphics[width=3.3in]{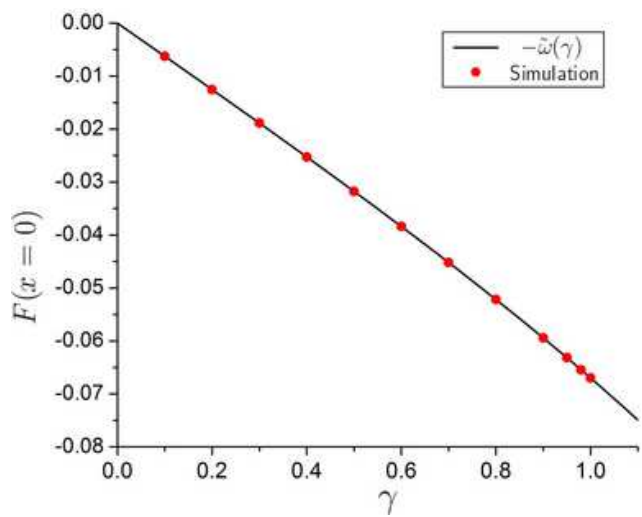}
\caption{\label{fig:zeroflux}  A plot of the  flux $F(x=0)$ at spin location $x=0$.  The measured error bars are smaller than the point size (errors of order $10^{-7}$) and are not included in the graph.}
\end{figure}

An easy initial check of our simulation is to plot the flux at the zeroth spin $F(x=0)$ versus the cold bath parameter $\gamma$.  We already argued in (\ref{eq:fluxat0}) that we  have the exact solution  $F(x=0) = -\tilde{\omega} =   \sqrt{4 \gamma^{-2}-1}- 2 \gamma^{-1}$ in units of $J$.  We see in figure \ref{fig:zeroflux} that the simulation results agree perfectly with this exact result for a wide range of $\gamma$.  Notice that as the cold bath temperature $T \rightarrow 0$, we have $\gamma \rightarrow 1$ and the flux increases in magnitude.  This makes sense as we increase the temperature gradient between the chains as we decrease $T$, forcing more energy through the junction between the chains.

\begin{figure}[!ht]
\centering
\includegraphics[width=3.3in]{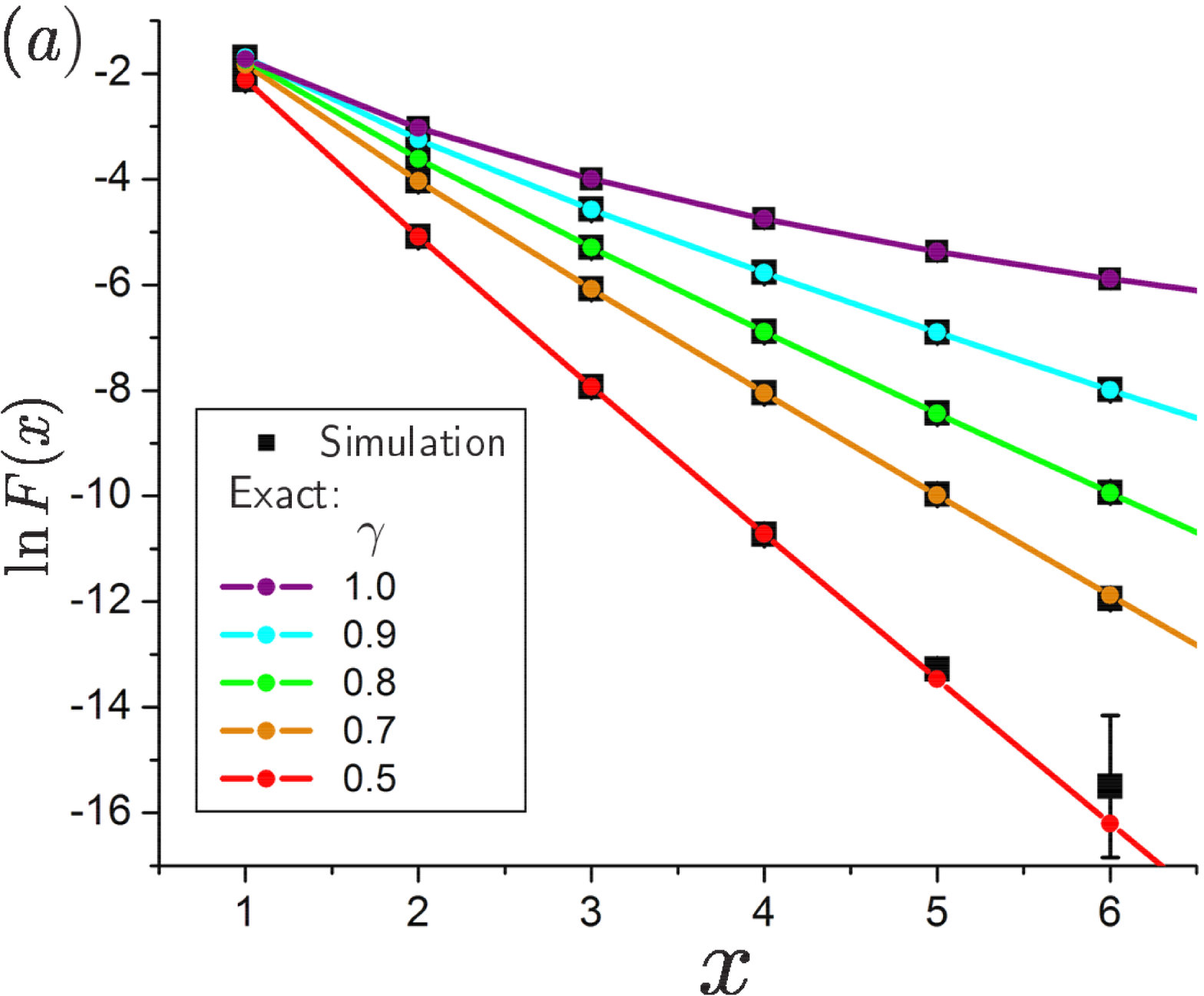}
\includegraphics[width=3.3in]{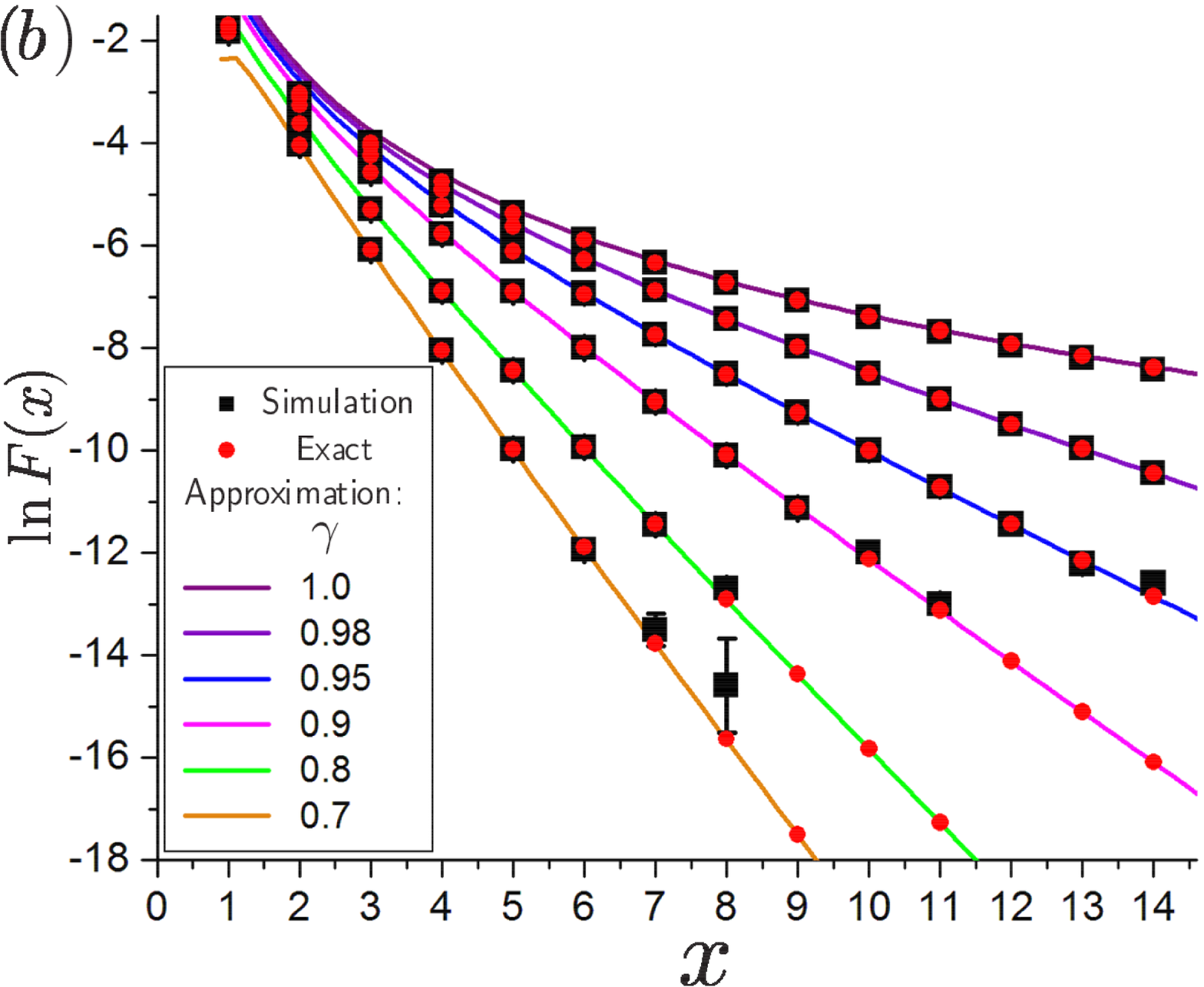}
\caption{\label{fig:totalfluxcomp}  
We compare the exact result in (\ref{eq:finalfexact}) with simulations for small values of the spin position $x$ in (a). The coloured lines in (a) are included to guide the eye.  In (b), we look at larger values of $x$ and show the result of the simulations (black squares), the exact solution (red disks), and also the Bessel function approximation (coloured lines) we derived in (\ref{eq:BesselKexpand}). The simulation results yielding flux values indistinguishable from zero are not included.}
\end{figure}
We can also test our exact result in (\ref{eq:finalfexact}) by numerically evaluating the integral in Mathematica.  The results are shown in figure \ref{fig:totalfluxcomp}(a).  The simulations again agree perfectly with the exact result for the fluxes $F(x)$ for all tested values of $x$ and $\gamma$.  Notice that for small values of $\gamma$ and large values of $x$, the magnitude of $F(x)$ is quite small and indistinguishable from zero in the simulations.  Thus, for these points we just show the numerically integrated exact result. Figure \ref{fig:totalfluxcomp}(b) also includes  approximation  (\ref{eq:BesselKexpand}).  Notice that the approximation agrees remarkably well with the simulation and exact result for small values of $x$ and captures the cross-over between the exponential and power-law decay of $F(x)$ as $\gamma \rightarrow 1$.

\begin{figure}[!ht]
\centering
\includegraphics[width=4in]{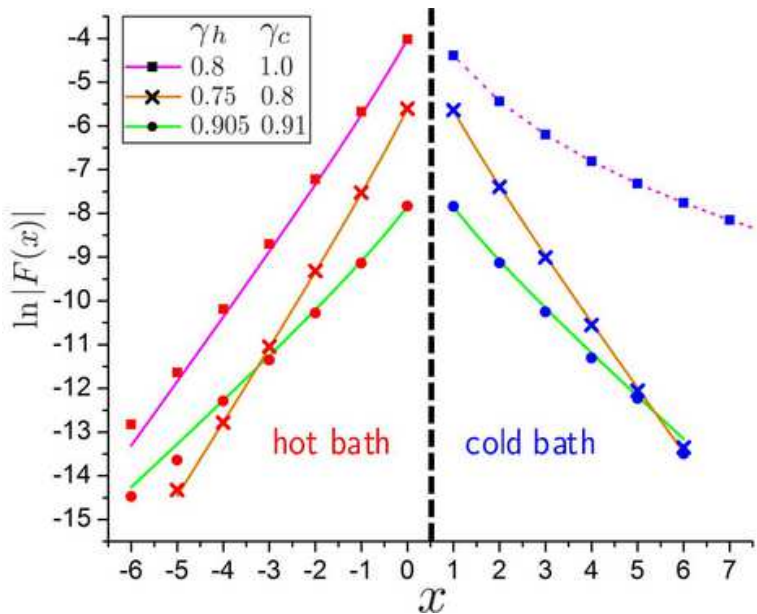}
\caption{\label{fig:2temp}  The plot markers are simulation results for $\ln |F(x)|$  for three separate chain simulations with different pairs of values for $\gamma_{h,c}$. The different marker styles are associated with different sets of coloured line segments, with colours representing the  $\gamma_{h,c}$ pairs shown in the plot legend.  The solid line segments are   best fits of the data in each bath for each simulated chain to the function $\ln |F(x)| = - | x  |/\xi_{h,c}- \ln |x|/2+b$, where $b$ is our fitting parameter and $\xi_{h,c}$ is given by (\ref{eq:corrlength}) with $\gamma = \gamma_{h,c}$. For spins with $x \leq 0$, we increment $x$ by 1 in the fitting function order to avoid the logarithmic singularity at $x = 0$.  The dashed magenta line simply connects the $\gamma_c = 1$ points as we do not have a conjectured fit function for this data. }
\end{figure}
Finally, simulations can be used as a tool to better understand the behavior of the model for arbitrary values of the bath temperatures $T_h$ and $T_c$ and to point us toward possible analytic solutions.  For example, we can check to see if the asymptotics we found in (\ref{eq:FLaplaceexpand}) for $T_c >0$ and $T_h \rightarrow \infty$ describe the flux decay when the hot bath is at a finite temperature. In  figure \ref{fig:2temp}, we fit simulation results for $\ln |F(x)|$ to the function $- |x|/\xi(T_{h,c})-(\ln |x|)/2+A$, where $\xi(T_{h,c})$ is half of the equilibrium correlation length for a chain at temperature $T_{h,c}$ and $A$ represents the fitting parameter and the overall amplitude of $|F(x)|$. Thus, we find that the simulations are consistent with  $|F(x)|$ decaying as $|F(x)| \approx A\,e^{-|x|/\xi(T_{h,c})}  /\sqrt{|x|}$ in \textit{both} baths.  This result confirms the intuitive picture of the flux in figure \ref{fig:setup}.  We see  that the flux in the hot bath must decay faster than in the cold one since we will have $\xi(T_h) < \xi(T_c)$.  An analytic check of this conjecture would be a natural extension of the results presented here.   In the $T_c \rightarrow 0$ ($\gamma_c \rightarrow 1$) limit, the flux decays more slowly into the cold bath, as shown by the points connected by the dashed line in figure \ref{fig:2temp}.  However, we seem to lose the $x^{-3}$ power law behavior for $T_h < \infty$ and find a different kind of long range decay.  We have no conjecture for the behavior of this decay, so no fit to the simulation data was attempted.  We also do not understand how the overall amplitude $A$ depends on $T_h$ and $T_c$. All we know is that it    should decay to zero as $T_h \rightarrow T_c$ and the whole chain approaches equilibrium.   This is evident in figure \ref{fig:2temp} where the chains with values of $\gamma_h$ close to $\gamma_c$  have smaller values of $|F(x)|$ at the junction around $x = 0$.

\section{ Conclusions and Outlook \label{sec:final}}

We have now examined in detail the exact results available for the energy flux through two Ising chains at different temperatures.   This flux is an interesting quantity as it is a purely non-equilibrium quantity which vanishes at equilibrium.  It is also physically relevant, as it describes the rate at which energy is transferred between the Ising chains due to a temperature gradient.  We found that when one of the chains is at infinite temperature ($T_h \rightarrow \infty$), an energy flux with magnitude $\tilde{\omega} = 2/ \gamma + \sqrt{4/\gamma^2-1}$ with $\gamma = \tanh(2 \beta J)$  is injected into the cold chain from the hot chain.  This energy then gets transported along the cold chain and eventually dissipates.  We have given an exact expression for this flux $F(x)$ and also an excellent approximation.  The main features of this flux is an exponential decay for $T_c>0$ and large $x$ with a decay length $\xi = -\{\ln \left[ \tanh(\beta J) \right] \}^{-1}/2$ equal to half of the correlation length of an infinite chain at equilibrium at temperature $T_c$.  As $T_c \rightarrow 0$, we get the maximum amount of energy injected into the cold chain and the flux $F(x)$ decays much more slowly, as the power law $x^{-3}$ at $T_c = 0$.  

Another way of understanding the  $T_c \rightarrow 0$ limit  is by focusing on the dynamics of the  boundaries between clusters of up and down spins.   We can treat these boundaries as particles by identifying any two spins that are anti-aligned.  For example, we can have spin configurations  $\uparrow \uparrow \circ \downarrow \downarrow$ or $\downarrow \downarrow \circ \uparrow \uparrow$, where we use the $\circ$ symbol to denote the boundary particle.  Notice that it costs no energy to flip either the second or third spin.  Thus, when either of these spins is selected in  a Monte Carlo simulation, it will flip with a probability of $1/2$ , moving the boundary either to the left or to the right.  Thus, these domain boundaries will perform unbiased random walks along the chain coupled to the cold bath with $T_c=0$ until they are removed at the two ends of the chain or until two domain boundaries annihilate via the transitions $\uparrow\circ \downarrow\circ \uparrow \rightarrow \uparrow \uparrow \uparrow$ and $\downarrow\circ \uparrow\circ \downarrow \rightarrow \downarrow \downarrow \downarrow$.    This pair-annihilation will release an energy of $4J$ into the cold bath.  Notice that the infinite temperature $x = 0$ spin will generate these particles at one end of the chain as it is allowed to flip even when its aligned with its neighbor at $x = 1$. Since this is the only mechanism by which we change the energy of the cold spin chain, when $\gamma = 1$, $F(x)$ measures the frequency of these annihilation events along the chain due to the injection of these boundary ``particles'' at $x = 0$. These kinds of particle models  (and their Ising model duals) have been the subject of much study \cite{farago1, farago2,  MZS2, MW} and many exact results for the steady state distributions and correlation functions are available via fermionic methods, Bethe ansatz techniques, etc.   This dual language might be useful in understanding what happens in the cold bath when $T_c = 0$ and $T_h < \infty$, as we were only able to fully understand what happens in the limit $T_h \rightarrow \infty$.  When $T_c > 0$, $F(x)$ also includes contributions from pair \textit{creation} of the boundary particles.

A natural generalization of the model considered in this paper to two dimensions is to look at spins on a square lattice. The lattice can  be partitioned into two halves, with spins in the two halves coupled to two different heat baths with temperatures $T_h$ and $T_c$. In two dimensions, there is a phase transition at finite temperature and the ordered phase (aligned spins) becomes stable at low temperatures.  This means that at low temperature, more energy is required to both form and move domain boundaries. The  boundaries at low temperature do not move as freely inside the system as they do in one dimension because  spins at  the boundaries will be typically surrounded by more aligned than anti-aligned spins.  Thus, we expect the fluctuations at the interface between the two halves of the lattice not to propagate very far at low temperatures.  The total energy flux between hot and cold baths coupled to spins on a two-dimensional lattice was studied previously \cite{Cor1, Cor2} to test the validity of fluctuation relations out of equilibrium.  It would be interesting to expand upon this research by looking at the spatial dependence of the flux.  For example, one could study the generalization of $F(x)$: the average rate of energy change due to spin flipping at a distance $x$ away from the interface.   

Preliminary simulation results on a square lattice  (with periodic BCs along the direction parallel to the interface between the two lattice halves and open BCs on the other two sides) suggest that, unlike in one dimension, the flux $F(x)$ is vanishingly small when $T_h = \infty$ and $T_c = 0$. The flux decays approximately exponentially for $T_h = \infty$ and arbitrary finite $T_c>0$.        We find that, assuming an exponential decay of $F(x)$, the associated decay length is small for all $T_c$ (on the order of the lattice spacing) and increases  as $T_c$ approaches the critical temperature $T_*\approx 2.3 J$ \cite{Pathria} for both $T_c < T_*$ and $T_c > T_*$.   Also unlike the one-dimensional case, the decay length of $F(x)$ does not appear to be simply related to the equilibrium correlation length. Of course, these results need to be checked with more careful simulations, especially when $T_c$ is close to $T_*$, where there might be novel critical behavior and associated power laws.   

 Our simulations were limited to small system sizes (squares of about 20 spins on a side) and we expect that there are finite-size effects.  A comprehensive account of these effects is beyond the scope of this paper, and the two-dimensional generalization of the model remains an interesting open problem for future study.  Such a study would require a detailed look at all the possible combinations of $T_h$ and $T_c$ relative to  $T_*$, as each half of the lattice is expected to behave qualitatively differently depending on whether its temperature and that of its partner is above or below the critical temperature.  As mentioned in the previous paper \cite{MLRKPZ2010}, we also expect a nontrivial magnetization profile at temperatures below the critical temperature.  Finally, the preliminary simulation results suggest that an interesting case is $T_h = \infty$ and $T_c = T_*$, where we seem to have the largest energy flux.

There are some more obvious open problems,  like the analytic behavior of $F(x)$ in one dimension for the case where  both $T_c$ and $T_h$ are finite.   We have formal expressions for this case, but we do not understand  precisely how $F(x)$ decays away from the junction for this more general case. We conjecture in section \ref{sec:simul} that for non-zero temperatures $F(x)$ decays exponentially into both baths. This should be checked  rigorously.  As mentioned in section \ref{sec:simul}, we also do not know the precise way in which $F(x) \rightarrow 0$ as the entire chain moves closer to equilibrium as $T_c \rightarrow T_h$.  Finally, as discussed in \cite{MLRKPZ2010}, there are countless ways to extend the calculations we presented here.  We could include different dynamics (such as Kawasaki spin-exchange), explore different boundary conditions, etc. 

It is interesting that the  far-from-equilibrium system studied here displays behavior reminiscent of a second order phase transition when  $T_h \rightarrow \infty$ and $T_c \rightarrow 0$:  We have a diverging decay length and a transition to a power law.  However, unlike an equilibrium phase transition, we do not have a general framework for understanding this crossover.  This means that exact results such as the ones presented in this work are especially useful in guiding more general studies
of non-equilibrium phenomena.

\ack This work was supported by the NSF Graduate Research Fellowship Program.  Simulations were performed on the Odyssey supercomputer at Harvard University.  The author warmly thanks R. K. P. Zia for inspiring this project and for his thoughtful guidance throughout, and E. S. Petrik for a careful reading of the manuscript and helpful discussions.

\end{document}